\documentclass[onecolumn]{aastex631}

\shorttitle{Pietras et al. 2022}
\shortauthors{}
\graphicspath{{./}{figures/}}

\usepackage{float} 
\usepackage{amssymb}
\usepackage{hyperref}

\begin{document}

\title{Statistical Analysis of Stellar Flares from the First Three Years of \textit{TESS} Observations}

\author[0000-0002-8581-9386]{M. Pietras}
\affiliation{Astronomical Institute, University of Wrocław, Kopernika 11, 51-622 Wrocław, Poland}

\author[0000-0003-1853-2809]{R. Falewicz}
\affiliation{Astronomical Institute, University of Wrocław, Kopernika 11, 51-622 Wrocław, Poland}
\affiliation{University of Wrocław, Centre of Scientific Excellence - Solar and Stellar Activity, Kopernika 11, 51-622 Wrocław, Poland}

\author[0000-0002-5006-5238]{M. Siarkowski}
\affiliation{Space Research Centre, Polish Academy of Sciences (CBK PAN), Bartycka 18A, 00-716 Warsaw, Poland}

\author[0000-0003-1419-2835]{K. Bicz}
\affiliation{Astronomical Institute, University of Wrocław, Kopernika 11, 51-622 Wrocław, Poland}

\author[0000-0001-8474-7694]{P. Pre\'s}
\affiliation{Astronomical Institute, University of Wrocław, Kopernika 11, 51-622 Wrocław, Poland}








\begin{abstract}

In this paper, we study stellar light curves from the \textit{TESS} satellite (Transiting Exoplanet Survey Satellite) for the presence of stellar flares. The main aim is to detect stellar flares using two-minutes cadence data and to perform statistical analysis. To find and analyze stellar flares we prepared automatic software WARPFINDER. We implemented three methods described in this paper: trend, difference, and profile fitting. Automated search for flares was accompanied by visual inspection. Using our software we analyzed two-minute cadence light curves of 330,000 stars located in the first 39 sectors of \textit{TESS} observations. As a result, we detected over 25,000 stars showing flare activity with the total number of more than 140,000 flares. This means that about 7.7\% of all the analyzed objects are flaring stars. The estimated flare energies range between $10^{31}$ and $10^{36}$ erg. We prepared a preliminary preview of the statistical distribution of parameters such as a flare duration, amplitudes and energy, and compared it with previous results. The relationship between stellar activity and its spectral type, temperature and mass was also statistically analyzed. Based on the scaling laws, we estimated the average values of the magnetic field strength and length of the flare loops. In our work, we used both single (about 60\%), and double (about 40\%) flare profiles to fit the observational data. The components of the double profile are supposed to be related to the direct heating of the photosphere by non-thermal electrons and back warming processes. 
\end{abstract}



\section{Introduction} \label{sec:intro}

  Many stars show flares similar to solar flares, especially the main-sequence stars of spectral types G, K, and M. This is most likely due to a similar flare mechanism - magnetic reconnection. These stars also have a convective zone, which for the coolest objects can even span the entire star \citep{2005nlds.book.....R}. The strength of the star's magnetic field is related to the motion of plasma due to convection \citep{doi:10.1146/annurev-astro-082708-101757}. 
  
  Stellar flares are highly energetic, rapid events, that occur during magnetic reconnection in the stellar coronae. Stored magnetic energy is impulsively released, converted into other forms of energy and consequently radiated across the entire electromagnetic spectrum from radio to gamma rays  \citep{2019ApJ...878..135K, 2011SoPh..271...57J}. 
  During the impulsive phase of solar flares and their analogs occurring on late spectral-type stars, beams of non-thermal electrons, accelerated somewhere in the solar or stellar coronae, stream along magnetic lines toward the chromosphere, where they heat plasma by colliding with the dense matter near the feet of the loops. The interactions of the non-thermal electrons with dense matter  also cause a strong and variable in time emission of  hard X-rays (HXR) in the feet of the loops (via the so-called bremsstrahlung process) \citep{1971SoPh...18..489B, 1978ApJ...224..241E}. The chromospheric matter, heated up to coronal temperatures, expands, filling the magnetic ropes (such a process is called chromospheric evaporation) and emits  soft X-rays (SXR) \citep{1984ApJ...287..917A, 1985ApJ...289..425F}. At the same time, a large amount of energy is radiated away from the loop-feet in the visible and ultraviolet significantly influencing the energy balance of the whole flare. The significant emission in the visual domain can be observed locally in solar flares (with high spatial and time resolutions), or it can be observed as a temporary variation of the integrated emission of the whole star.

The optical emission of a solar flare consists of a continuous spectrum (continuum) and chromospheric lines. During the majority of solar flares, the intensity of the continuum does not increase noticeably, while the intensity of spectral lines, formed mainly in the solar chromosphere and the transition region (TR), may increase significantly \citep{2015SoPh..290.3487K}. The continuum optical emission is formed mainly in the solar/stellar photosphere, where the temperature and density are almost undisturbed during flares, and therefore the intensity of continuum is not affected by the flares. On the contrary, the hydrogen, calcium, magnesium, and other chromospheric lines are formed in the middle or higher chromosphere \citep{1978stat.book.....M}, where the plasma parameters are strongly modified by different flare heating mechanisms. Therefore, the emission in these lines is strongly enhanced during flares. The main plasma parameters, which may influence the intensity of the spectral lines, are the temperature and density of the plasma. During solar flares, the values of both parameters increases due to non-thermal plasma heating and evaporation of the chromospheric plasma. The amount of such increases depends on the energy budget of all heating and cooling mechanisms.
	
The above-described situation is valid for most average-size solar flares. However, in some big flares the flux and the energy of accelerated non-thermal particles are as height that the heating of the low-lying solar photosphere or the lower chromosphere is possible. Then, the continuum optical emission may arise from some flare areas. Such flares are called “white-light” flares and they were observed in the past  \citep{1974SoPh...38..499M,1982SoPh...80..113H,1984SoPh...92..217N}.

 Assuming the photospheric origin of white-light flare emission, the most probable mechanism that may be responsible for this continuum is H$^-$ emission \citep{1994ApJ...429..890D}. It is possible that the photosphere is not heated directly by non-thermal particles but during the process of “radiative backwarming,” where the non-thermal electrons deposit their energy in the chromosphere and then, the hotter chromosphere radially heats the lower-lying photosphere \citep{1995A&AS..110...99F}. More recent studies show that the flare continuum emission can be formed in the thin “layer” of the chromosphere in the process of hydrogen recombination \citep{potts2010}.

  Research and observations of white-light flares from nearby stars have been rapidly developing in recent years thanks to large optical photometric surveys. For example, \citet{Davenport_2014}, \citet{2011AJ....141...50W}, and \citet{2014ApJ...797..121H} used data from the \textit{Kepler} mission, for which time resolution was 1 minute (SC - short cadence) or 30 minutes (LC - long cadence). Other examples are Sloan Digital Sky Survey (SDSS) \citep{2000AJ....120.1579Y}, EVRYSCOPE \citep{2014SPIE.9145E..0ZL}, All-Sky Automated Survey for Supernovae (ASASSN) \citep{2016RMxAC..48...78B} or Next Generation Transit Survey (NGTS) \citep{2018MNRAS.475.4476W}.
  
  Duration times of white-light flares usually vary from minutes to hours, with a fast rise and a slower exponential decay profiles. Typical flare energies range from 10$^{26}$ to 10$^{32}$ erg. So-called super-flares with energies between 10$^{32}$ and 10$^{36}$ erg are also observed \citep{2000ApJ...529.1026S,2013ApJS..209....5S}. Flare activity is thought to be correlated with fast rotation and a late spectral type of a star. About 40\% of M-type dwarfs are flaring stars \citep{2017ApJ...849...36Y,gunther,2019ApJ...881....9H}. Moreover, detecting flares on late-type stars is easier because of the high flare contrast caused by a low surface temperature. The \textit{TESS} mission provides us with the opportunity to study a large number and a variety of stellar flares. 
  
   In this paper, we examined light curves from the \textit{TESS} satellite (Transiting Exoplanet Survey Satellite) for the presence of stellar flares. We analyzed stellar flares from the first 39 sectors of the \textit{TESS} mission. Our study investigates the preliminary statistics of stellar flares. It also includes relationships between stellar flares and such stellar properties as mass, effective temperature, and spectral type. In Section \ref{sec:observations} we describe the \textit{TESS} observations. Our software and methodology to find flare candidates are detailed in Section \ref{sec:software}. The results are presented and compared with previous findings in Section \ref{sec:results}, and a discussion is provided in Section\ref{sec:discussion}.

  \section{Observations}  \label{sec:observations}

  \textit{TESS} \citep{2014SPIE.9143E..20R} is a space-based telescope launched in April 2018. Its orbit is highly elliptical, with a period of 13.7 days. The main goal of the mission is to search for exoplanets using the transit method. \textit{TESS} instrument consists of four 10.5-cm telescopes, each covering a field-of-view of 24\,$\times$\,24 degrees. In a single pointing \textit{TESS} cameras cover a field of view of 24\,$\times$\,96 degrees. A given sector is observed through two satellite orbits, that is, for slightly less than a month. The first part of the mission, completed in July 2020, consisted of observations of 26 sectors in both hemispheres covering about 85\% of the sky. During the ongoing extended part of the mission, observations of the next 29 sectors (27 -- 55) are carried out. The sectors partly overlap, which results in the presence of the continuous viewing zone in the vicinity of both ecliptic poles. During the primary part of the mission, two cadences were realized, 2-min cadence for a sample pre-selected objects and 30-min cadence for the whole covered FoV. The latter cadence is offered as Full Frame Images that can be used to extract photometry for any object in the FoV. During the extended part of the mission, two short-time cadences, 2-min and 20-s, were chosen for selected objects, while FFIs are offered with a 10-min cadence. The FFIs have angular resolution of 21 arcsec per pixel.
  
 Almost 330 000 stars observed with TESS in two-min cadence are in our sample, therefore we can study a huge number of stellar flares with a high signal-to-noise ratio. We selected calibrated, short cadence PDCSAP (Pre-search Data Conditioning Simple Aperture Photometry) light curves downloaded from the Mikulski Archive for Space Telescopes (MAST). Moreover, we required the Quality flag to be equal to 0 and normalized light curves by dividing the flux by the mean flux of a star. If available we took mass, radius, spectral types, and other stellar parameters from MAST and SIMBAD databases if possible. All the stars with spectral types earlier than F1 were rejected from our analysis. The effective temperatures of the flaring stars in our sample is limited to stars cooler than about 8000 K with masses smaller than about 1.7 M$_\odot$. This limitation is enforced by numerous rapid changes in light curves of hot stars, which are similar in shape to stellar flares. These changes were often misclassified by our algorithm as flares. Stellar variability of the RR Lyrae, $\delta$ Scuti, $\gamma$ Doradus, and SX Phoenicis was particularly  problematic. After analyzing stars from the first 39 sectors of \textit{TESS} observations, we found more than 25,000 flaring stars and 140,000 individual events.

  \section{Software}  \label{sec:software}

The software WARPFINDER (Wroclaw AlgoRithm Prepared For detectINg anD analyzing stEllar flaRes) prepared by us is written in the IDL (Interactive Data Language) and Python programming languages, which are mainly used for stellar flare identification and analysis. The software has also been adapted to automatically download information about stars from MAST and SIMBAD (Set of Identifications, Measurements and Bibliography for Astronomical Data), as well as data from the \textit{TESS} satellite. Our automatic procedure also rejects most false detections such as asteroids, transits, or stellar pulsations. The appearance of an asteroid is usually observed in a stellar light curve as a strong, symmetrical increase of flux. Our software checks the asymmetry of each detection and rejects overly symmetrical phenomena. In addition, this shape is not well described by the flare profile we assumed. Many false detections occur during transits when the flux increases rapidly. To eliminate them, we added to the software the conditions regarding the relationship between the value of the signal at the beginning, at the maximum, and at the end of the flare. The most problematic, however, was the numerous false detections associated with short-period stellar pulsations. For this reason, we have limited our analysis to only stars with spectral types later than F0. The remaining false detections are rejected during partial visual inspection. The tables and diagrams containing data about detected stellar flares and the flaring stars are the final results of using software. 
\subsection{Methods}

We prepared an automated, unsupervised three-step method for finding flare events from \textit{TESS} data.  To search for potential flaring stars, we analyzed the 307 281 available targets with \textit{TESS} PDCSAP light curves from the first 39 sectors. In our research we need data without systematic trends, so PDCSAP flux was more appropriate than SAP (Simple Aperture Photometery) flux. In this work, we used light curves with a time resolution of two minutes.

\subsubsection{Trend method}

One of the methods of finding stellar flares is the trend method. The first step is based on consecutive de-trending of light curves using smoothing with several window lengths. This method was inspired by automated procedures described by \citet{Davenport_2014} and tested on \textit{Kepler} data. We smoothed the observed flux with the running average (smooth) function with a window length of 175 points (350 min). For this calculated trend we determined the standard deviation, and then calculated the new trend, rejecting all points protruding more than $ 3 \sigma$ above the trend. We repeated this step recursively five times. For further analysis, we selected all points $1 \sigma$ above the trend. We defined all the series of these points where at least four points protrude above the $2.5 \sigma$ level as flares. They do not have to be consecutive points. There can be individual points between them where the deviation is less than  $2.5 \sigma$, but greater than $1 \sigma$. The probability that a single point occurring randomly above the $2.5 \sigma$ threshold is 0.0062. The likelihood of randomly getting  4 out of 7 such consecutive points drops to about $5 \times 10^{-12}$. This gives us a chance for one false detection of a flare in one sector of properly detrended observations around $1 \times 10^{-7}$.

We repeat above operation for the next five smoothing windows with lengths of 31, 75, 121, 221, and 311 points, respectively. The next flares determined in this way are added to the previous ones if the times of their maximums do not lie within the limits of the flares determined earlier. We have applied a few different values of smoothing windows because many flaring stars have more than one period of variability. These windows can be chosen as free parameters in our code and the values used here were selected experimentally on the data from TESS sector 1, 2 and 3. We checked these values for another set of thousands of flares and they seem to work pretty well for data with a 2 minute cadence.

There are many false detections caused by non-adequate  (usually too long)  trend windows. Before the visual inspection, we verify these bad signals using other search methods in the next two steps of our automatic procedure.

\subsubsection{Difference method}

The second method of finding stellar flares is the difference method whose idea was taken from the paper by \citet{2013ApJS..209....5S}. The method is based on checking the flux difference between two consecutive points. In our method, contrary to the authors of the cited paper, we do not check the percentage distribution of differences, but the standard deviation instead. We assumed the study of points whose sigma is greater than $3 \sigma$ as a default value. The spread of points is analyzed only for the positive values. All points that exceed $3 \sigma$ are marked as potential flare detection and the times are compared with the hits detected by the trend method. It should be noted, that only the mutual detection of probable flares in a designated time is treated by the software as true detection and passed on to the third level of verification.

\subsection{Flare profile method}

The prepared list of potential stellar flares, which were detected by both previous methods, is then verified by a detailed analysis of their light curve profiles. In the first step of the profile method, we try to properly determine the beginning and the end time of a stellar flare. From the trend method, we obtain an estimated range of the flare duration. We use it to find the pre-flare and post-flare background approximated by a linear function. A linear background is sufficient in most cases. The standard duration of the flares usually does not exceed several dozen minutes. The percentage of stars rotating fast enough to affect the background determination is negligible in such a large sample of detected flares. Then we iteratively determine a new time of the start and the end of the flare and repeat the whole procedure 10 times. This value was selected on the basis of tests performed by us. In each iteration, we try to fit the profile for each test start and end point and choose the best fit based on the ${\chi}^2$ value. Further, if the best solution is correct, we detected it as a flare. The final linear trend is fitted to the observational points and outliers are treated as a stellar flare. 

In the next step, we fit the assumed flare profiles to the observational data. The first profile (Equation \ref{eq:profile1}) of a flare is given by equation \citep{2017SoPh..292...77G}:

\begin{equation}\label{eq:profile1}
f(t) = \frac{1}{2} \sqrt{\pi}A C  \cdot exp[D (B-t) + \frac{C^2D^2}{4}] [erf(\frac{2B+C^2D}{2C})-erf(\frac{2B+C^2D}{2C}-\frac{t}{C})]
\end{equation}

where A, B, C, D are parameters, \textit{t} is the time and \textit{erf} is the error function. This profile is a convolution of the Gaussian profile (heating function) and exponential decay (function describing processes of energy loss). The A parameter is related to the amplitude of the flare. Parameter B determines the maximum energy release. Parameter C informs about the time scale of the energy release. The inverse of the D parameter is related to the flare decay time. We noticed that some of the flares do not fit well with only one profile. For this reason, we also performed a fit created from two similar profiles. In one case, we used profiles characterized by a small shift in the time of the maximum of both components (Equation \ref{eq:profile2_1B}). These components have the same value as the B parameter. In the second case, we allowed both components to have large shifts between the time of their maximums, and thus different values of the B parameter (Equation \ref{eq:profile2_2B}).

\begin{equation}\label{eq:profile2_1B}
   f(t) = \int_{0}^{t}(A_1 \cdot exp[\frac{-(x-B)^2}{C_1^2}] \cdot exp[-D_1(t-x)] + A_2 \cdot exp[\frac{-(x-B)^2}{C_2^2}] \cdot exp[-D_2(t-x)])dx    
\end{equation}

\begin{equation}\label{eq:profile2_2B}
   f(t) = \int_{0}^{t}(A_1 \cdot exp[\frac{-(x-B_1)^2}{C_1^2}] \cdot exp[-D_1(t-x)] + A_2 \cdot exp[\frac{-(x-B_2)^2}{C_2^2}] \cdot exp[-D_2(t-x)])dx    
\end{equation}

We checked the quality of each of the flare profile fits. We used the ${\chi}^2$ statistic for this purpose. The fit with has the lowest ${\chi}^2$ value was selected for a further analysis. For good fits, this value usually does not exceed 3.0 of reduced ${\chi}^2$ (taking into account the number of data points and the number of degrees of freedom). To distinguish a stellar flare from data noise we also used the probability density function of F-distribution. We calculate the reduced ${\chi}^2$ for the observational data with the fitted flare profile and the estimated trend. Next, we computed the cumulative distribution function for an F-distribution with defined degrees of freedom. We accepted the detection as a flare if the probability is less than the cutoff value. After carrying out the earlier tests, we assumed that this value cannot exceed 0.1. Additional methods to reject false detections are: the calculation of the flare profile, skewness and the bisectors of the profile. For the purposes of this work, we define skewness in general as a measure of flare profile asymmetry. We assumed that a stellar flare should have a shorter rise time than its decay time. For this reason, the skewness of the profile should be a positive value. The bisector method gives information about the symmetry of the flare profile. Bisectors are defined by lines at the level of 10\%, 20\%, etc. of the maximum signal value. If the bisector method indicates an overly symmetrical profile (a value of about 0 or less), then the detection must be verified visually by a software user. Moreover, detections with a duration of less than 12 minutes (six points) are rejected.

After the analysis, separate data sets (three folders) with detections are created. The first one contains mostly real flares. The second set is for uncertain events that require visual inspection.  The third set contains the detections rejected by the software.

\subsection{A flare injection-recovery tests} \label{subsec:detection_tests}
A flare injection-recovery test were performed to validate the results found via all three methods described in this paper. As in \citet{gunther}, we decided to use two separate tests. For each of them, we selected randomly 1000 light curves of spectral type F, G or K stars, as well as early M dwarfs and late M dwarfs. For the first test, we artificially added 10 time-separated flares to each light curve. In the second test, we also inserted 10 flares, but they were only separated by 10 minutes up to a maximum of 1 hour. Each of the inserted flares was described by a single profile described above with randomly selected parameters. The range of this parameters was determined on the basis of the stellar flares we have observed. 

\begin{figure}[H]
    \centering
    \includegraphics[width=0.83\textwidth]{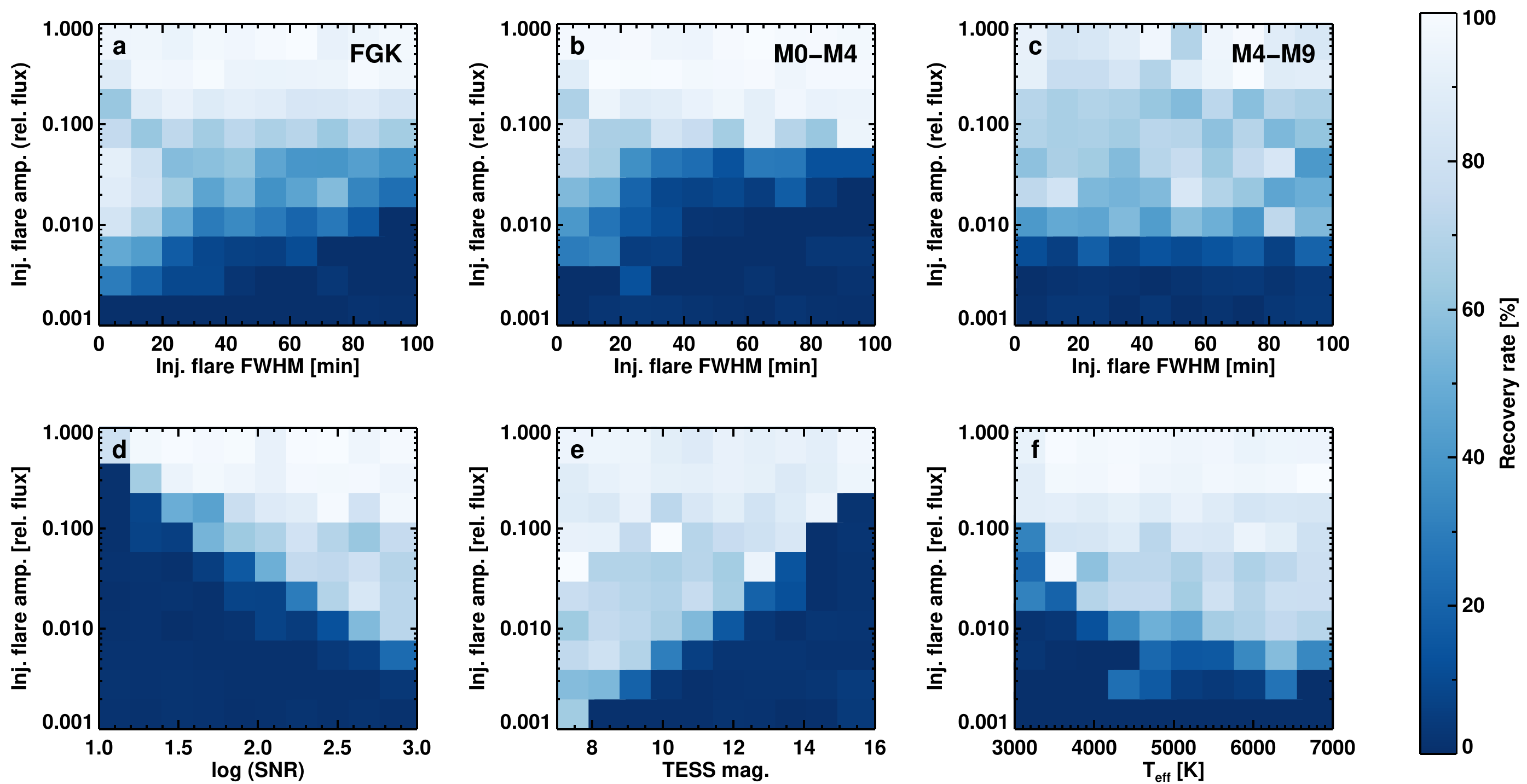}
    \caption{The results of a flare injection-recovery tests. The 
three top panels show the recovery rate as a function of the injected flares’ amplitudes and FWHM in minutes. From the left, these are the results for dwarfs of the F, K, or G (a), M0-4 (b) and M4-9 (C) types.  The three bottom panels show the recovery rate as a function of the injected flares’ amplitudes and the logarithm of SNR (Signal-to-Noise Ratio) of a stellar lightcurve (d), \textit{TESS} magnitude (e), and stellar effective temperature (f). On the right, there is the colorbar of recovery rate in percentage.}
    \label{fig:injection}
\end{figure}

After proper preparation of the light curves, we tested our three-step software. In the case of both tests, we considered the flares as detected if their maximum was determined within the previously assumed duration. The Figure \ref{fig:injection} shows the test results we obtained. From the top three panels it can be seen that well detected flares usually have amplitudes greater than 0.01. Moreover, for M-type dwarfs, the mean recovery rate is lower. This is due to the lower brightness and SNR (Signal-to-Noise Ratio) of lightcurves of these stars. These dependencies are well illustrated by the three lower panels. It can be noticed that the recovery rate significantly decreases with the SNR, \textit{TESS} magnitude and effective temperature of stars.

\subsection{Flare energy}  \label{subsec:flareenergy}

The maximum luminosity and the total energy of flares are estimated in two different methods. The first one requires a flare amplitude, a flare duration, a stellar luminosity, and a radius. \citet{2016Kowalski} showed using hydrodynamic simulations that a temperature of about 10000 K is needed in order to correctly reproduce the flare emission on M stars in the range of white light. Thus, we assumed black body-radiation and the effective temperature of the flare ($T_{flare}$) about 10000 K \citep{2013ApJS..209....5S,1980ApJ...239L..27M,1992ApJS...78..565H}. 

The flare amplitude ($C'_{flare}$) is defined as follows:

\begin{equation}
C'_{flare}= F'_{flare}/F'_{star}
\end{equation}

where $F'_{star}$ and $F'_{flare}$ are observed flux of the star and the flare. $A_{flare}$ is the flare area.

\begin{equation}
F'_{star}=\pi R^2_{star} \int S_{\rm{TESS}} B_{\lambda (T_{eff})} d\lambda 
\end{equation}
\begin{equation}
F'_{flare}=A_{flare}\int S_{\rm{TESS}} B_{\lambda (T_{flare})} d\lambda 
\end{equation}

\begin{equation}
A_{flare}=C'_{flare} \; \pi R_{star}^2 \frac{\int S_{\rm{TESS}} B_{\lambda (T_{eff})} d\lambda}{\int S_{\rm{TESS}} B_{\lambda (T_{flare})} d\lambda}
\label{eq:area}
\end{equation}

where
$S_{\rm{TESS}}$ is the \textit{TESS} response function, $B_{\lambda}$ is the Planck function. The \textit{TESS} response function was defined as in \citet{2014SPIE.9143E..20R}.

The bolometric flare luminosity ($L_{flare}$) and the energy ($E_{flare}$) could be calculated from:

\begin{equation}
L_{flare}=\sigma_{SB} T_{flare}^4 A_{flare}
\end{equation}

\begin{equation}
E_{flare}=\int _{t1}^{t2} L_{flare} dt
\end{equation}

where $\sigma_{SB}$ is the Stefan–Boltzmann constant.

For a star for which information about radius is available, it is possible to estimate the luminosity and energy emitted during the flare. According to \citet{2013ApJS..209....5S}, the total energy uncertainty in this method is about 60\%.

To estimate the energy of detected flares in the \textit{TESS} detector bandpass we also used the modified method proposed by \citet{2007AN....328..904K}. It was described in \cite{2018A&A...616A.163V} and \cite{2019ApJ...884..160V}. This method is based on the following integrating normalized flare intensity with subtracted background during the flare event:

\begin{equation}
\varepsilon_{\rm{TESS}} = \int _{t1}^{t2} C'_{\rm{flare}} dt
\end{equation}

\noindent where $t_1$ and $t_2$ are the times of a start and an end of the flare. The integral $\varepsilon_{\rm{TESS}}$ is relative flare energy. Next, we had to estimate flux of the star in a selected interval of wavelengths ($\lambda_1,\lambda_2$), which are wavelengths of the \textit{TESS} detector bandpass. To estimate flux of a star we multiply spectrum of the star $\mathcal{F}(\lambda)$, taken from ATLAS9\footnote{https://wwwuser.oats.inaf.it/castelli, see also \cite{2003IAUS..210P.A20C}.} for stars $\log(g)$, $T_{\rm{eff}}$ (in spectra we assumed metallicity of the Sun and $v_{\rm{turb}} = 2$ km s$^{-1}$), with \textit{TESS} response function $S_{\rm{TESS}}(\lambda)$ multiplied by stars area with radius $R$. Due to the use of the theoretical spectrum of stars from ATLAS9, there is no factor 4 in the following equation:

\begin{equation}
F_{\rm{star}} = \pi R_{star}^2 \int _{\lambda_1}^{\lambda_2} \mathcal{F}(\lambda)S_{\rm{TESS}}(\lambda)  d\lambda 
\end{equation}

\noindent To estimate the flare energy in the \textit{TESS} detector bandpass $E_{\rm{flare}}$, we multiplied star's flux $F_{\rm{star}}$ in a selected interval of wavelengths by relative flare energy $\varepsilon_{\rm{TESS}}$:

\begin{equation}
E_{\rm{flare}} = F_{\rm{star}} \cdot \varepsilon_{\rm{TESS}}
\end{equation}

Flare energies estimated in this way are not bolometric. For this reason, they have lower values than for the first method, which is based on \citet{2013ApJS..209....5S}.

\subsection{Problematic cases}

Our program for automatic detection of stellar flares WARPFINDER may not work properly in a few cases. We observed a particularly large number of false detections on stars with spectral types earlier than F0. These are stars that are often characterized by a rapid variation in the light curve. An example is the RR Lyrae stars, whose periods of variation can be several hours. Moreover, the shape of the brightness changes is similar to the flare profile assumed by us. It is characterized by a rapid growth phase and slow decay. Similar variability, classified as flares, is observed for $\delta$ Scuti, $\gamma$ Doradus, and SX Phoenicis stars.

Flare detection is also difficult for very cool M spectral type stars. Due to their low brightness, light curves have a low signal-to-noise ratio. Therefore, faint flares are not detectable by our software. This conclusion is also confirmed by injection-recovery tests performed by us.

Moreover, our flare catalog may include some false detections that are only observed once per star in all available sectors. This is due to a random spike in the signal accidentally classified as a flare. However, we decided not to exclude such stars from the analysis, so as not to remove information about real flaring stars.

\section{Results}  \label{sec:results}

\begin{figure}[h]
    \centering
    \includegraphics[width=0.9\textwidth]{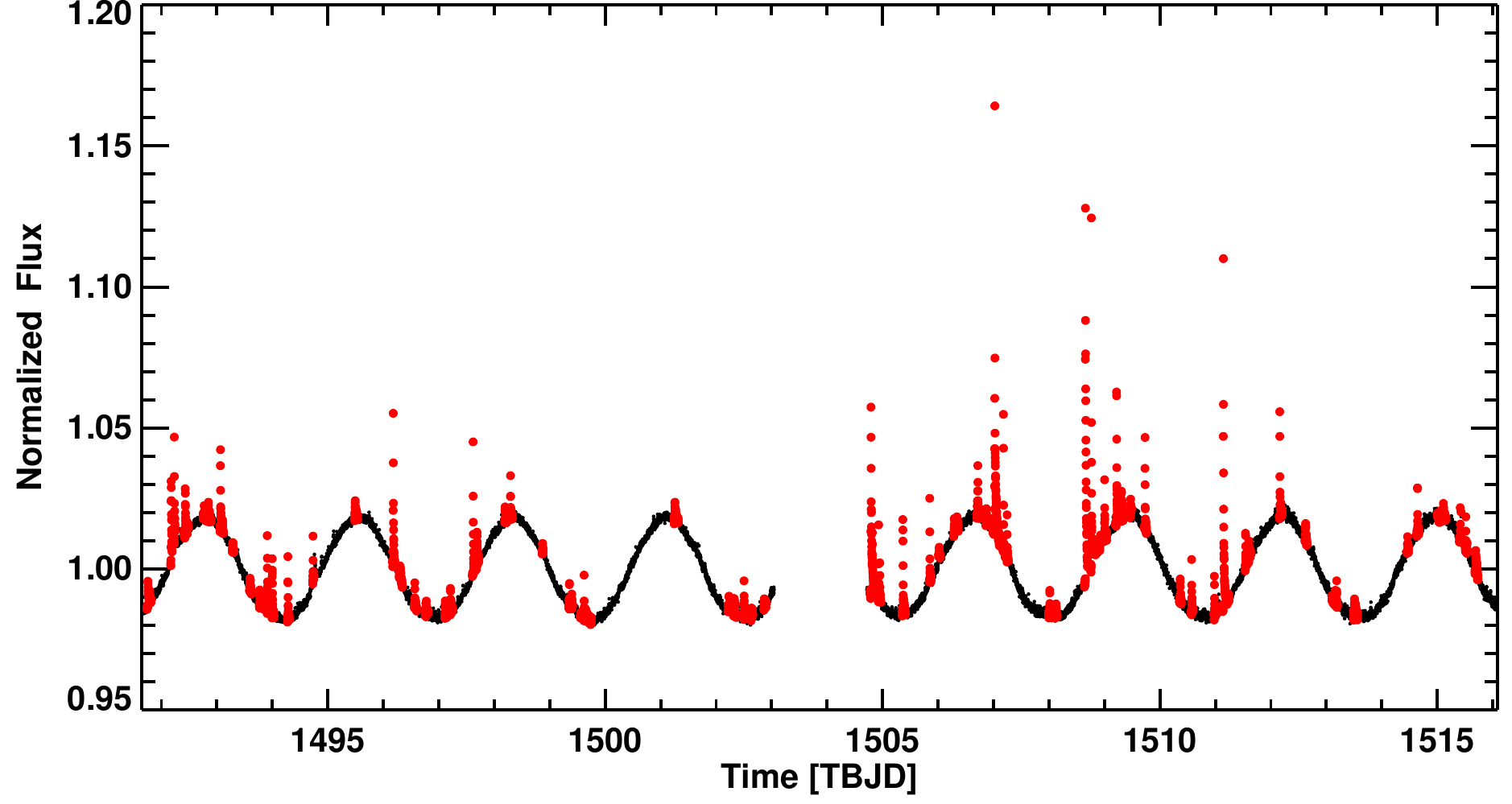}
    \caption{The light curve of the TIC266744225 (YZ CMi, sector 7). The flares found by full WARPFINDER pipeline are marked with red dots.}
    \label{fig:YZ_CMi}
\end{figure}

In this paper, we presented the results of the analysis of 329,176 stars from the first 39 sectors of \textit{TESS} observations. We found 25,229 flaring stars and 147,683 individual flares. The catalog with all detected flares is available in machine-readable form. Approximately 7.7\% of all observed stars show flaring activity. This value varies depending on a spectral type of the studied stars and is greater than 50\% for the stars of spectral type M. On average, there are four flares per each flaring star observed during about 25 days (one \textit{TESS} sector). We observed up to 69 flares on one star in one sector (TIC266744225, Figure \ref{fig:YZ_CMi}). Moreover, on the TIC150359500 we found 863 events during 23 sectors of observation. 

\begin{figure}[H]
    \centering
    \includegraphics[width=0.6\textwidth]{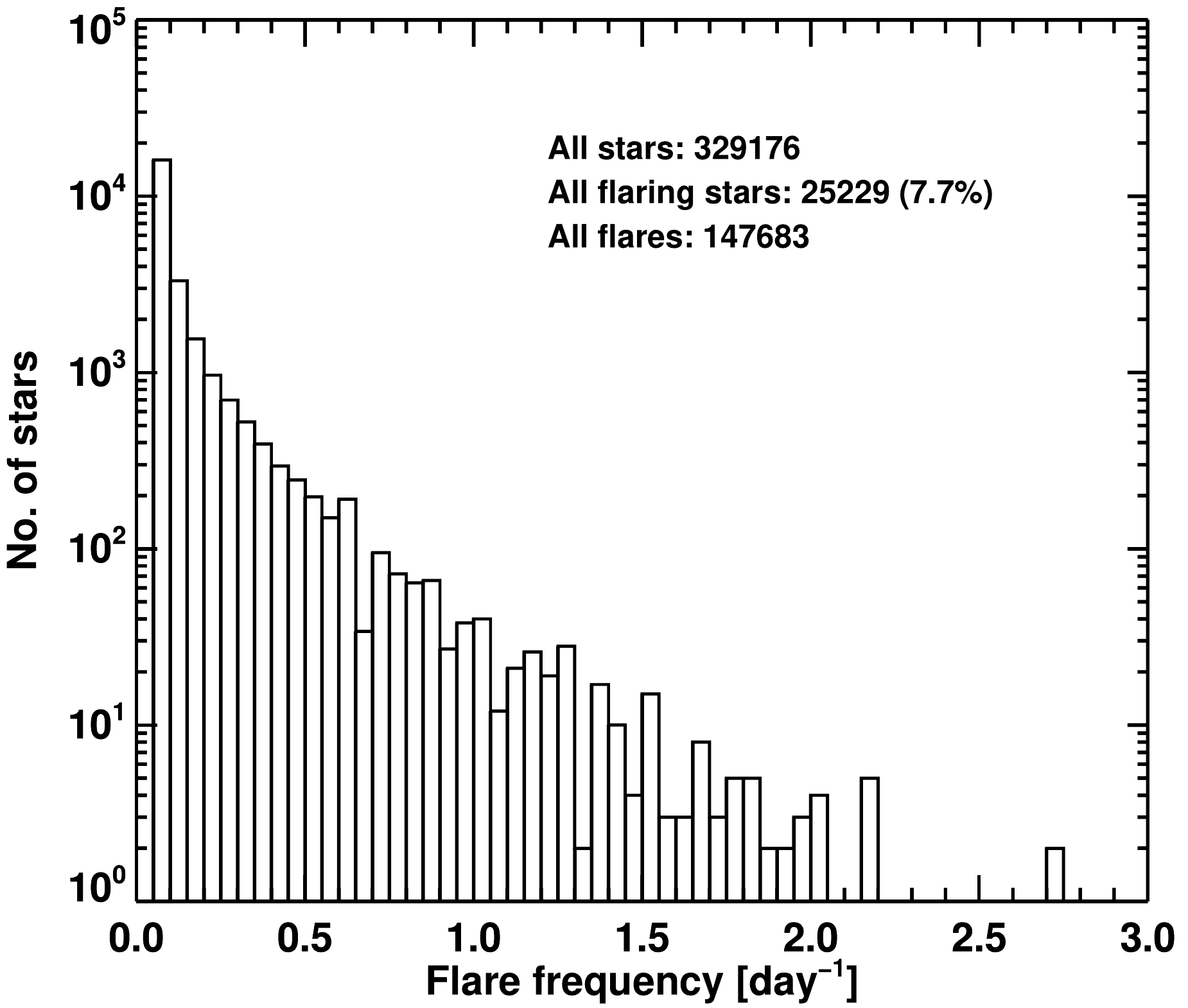}
    \caption{The histogram of flares' occurrence frequency per day.}
    \label{fig:flares_stat}
\end{figure}

Figure \ref{fig:flares_stat} shows the histogram of flares'f occurrence frequency per day. In most cases,  one flare is observed on one star every ten days. There are, however, a few very active stars, where average flare activity equals more than two events a day (TIC266744225, TIC142206123, TIC441420236). A similar result was obtained by \citet{2016ApJ...829...23D}, but for a much larger sample of flares. In  Table \ref{tab:table_sectors} we compare the number of stars observed in each of the 39 sectors with the number of flaring stars and individual flares. The hypothesis about the uniform distribution of stellar flares in sectors was rejected based on the performed Pearson's ${\chi}^2$ test. Pearson's ${\chi}^2$ test against the hypothesis of uniform distribution returns a value of 840 for 35 degrees of freedom, which exceeds the acceptable value for the assumed probability of the hypothesis. Moreover, there is an increase in the percentage of flaring stars during the \textit{TESS} Extended Mission (sectors above 27). This may be related to the change in the way of selecting observation targets. Now, about ~80\% of targets per sector come from Guest Investigator (GI) programs. In total, we found 9480 flaring stars during the first year of \textit{TESS} observations (S01-S13, ecliptic declination from -90 $^\circ$ to -54 $^\circ$). For comparison, during the third year of the mission, when the southern hemisphere was re-observed (S27-S39), we found 10 661 flaring stars, including 8913 new objects.

\newpage

\begin{deluxetable}{ccccc}[H]
\tablenum{1}
\tablecaption{The number of all stars, the flaring stars, the ratio of flaring stars to all stars and the individual flares observed in each sector
\label{tab:table_sectors}}
\tablewidth{0pt}
\tablehead{
\colhead{Sector} & \colhead{All stars} &  \colhead{Flaring stars} &
\colhead{ratio {[}\%{]}} & \colhead{Flares}}
\startdata
01     & 16000     & 990          & 6.2           & 3661              \\ 
02     & 16000     & 1043          & 6.5           & 4564              \\ 
03     & 16000     & 796           & 5.0           & 3084              \\
04     & 20000     & 1525          & 7.6          & 4567               \\ 
05     & 20000     & 1356          & 6.8           & 5760              \\ 
06     & 20000     & 1076          & 5.4           & 3719              \\ 
07     & 20000     & 924          & 4.6           & 3079              \\ 
08     & 20000     & 950          & 4.8           & 2815              \\ 
09     & 20000     & 943          & 4.7           & 3167              \\ 
10     & 20000     & 1072          & 5.4           & 3849              \\ 
11     & 20000     & 1063          & 5.3           & 3674              \\ 
12     & 20000     & 1055          & 5.3           & 3383              \\ 
13     & 20000     & 930          & 4.7           & 3333              \\ 
14     & 20000     & 1088          & 5.4           & 3096              \\ 
15     & 20000     & 958          & 4.8           & 2778              \\ 
16     & 20000     & 968          & 4.8           & 2851              \\ 
17     & 20000     & 944          & 4.7           & 2422              \\ 
18     & 20000     & 1058          & 5.3           & 3276              \\ 
19     & 20000     & 1025          & 5.1           & 3449              \\ 
20     & 20000     & 1235          & 6.2           & 3929              \\ 
21     & 20000     & 1221          & 6.1           & 3910              \\
22     & 20000     & 1064          & 5.3           & 3392              \\ 
23     & 19999     & 1152          & 5.1           & 2991              \\ 
24     & 20000     & 1094          & 5.5           & 3436              \\ 
25     & 20000     & 1129          & 5.7           & 3354              \\ 
26     & 20000     & 1140          & 5.7           & 3291              \\ 
27     & 19999     & 954          & 4.8           & 3452              \\
28     & 20000     & 916          & 4.6           & 3287              \\ 
29     & 19999     & 1182          & 5.9           & 4307              \\ 
30     & 19999     & 1148          & 5.7           & 4569              \\ \
31     & 19999     & 1236          & 6.2           & 4575              \\ 
32     & 19995     & 1539          & 7.7           & 6274              \\ 
33     & 19991     & 1163          & 5.8           & 4575              \\ 
34     & 20000     & 1359          & 6.8           & 6274             \\ 
35     & 20000     & 1382          & 6.9           & 3717              \\ 
36     & 20000     & 1254          & 6.3           & 4343              \\ 
37     & 20000     & 1334         & 6.7           & 4798              \\ 
38     & 20000     & 1516         & 7.6           & 5280              \\ 
39     & 20000     & 1334         & 6.7           & 4793              \\ 
\enddata
\end{deluxetable}

\subsection{Parameters of flaring stars}

\begin{figure}[H]
\gridline{\fig{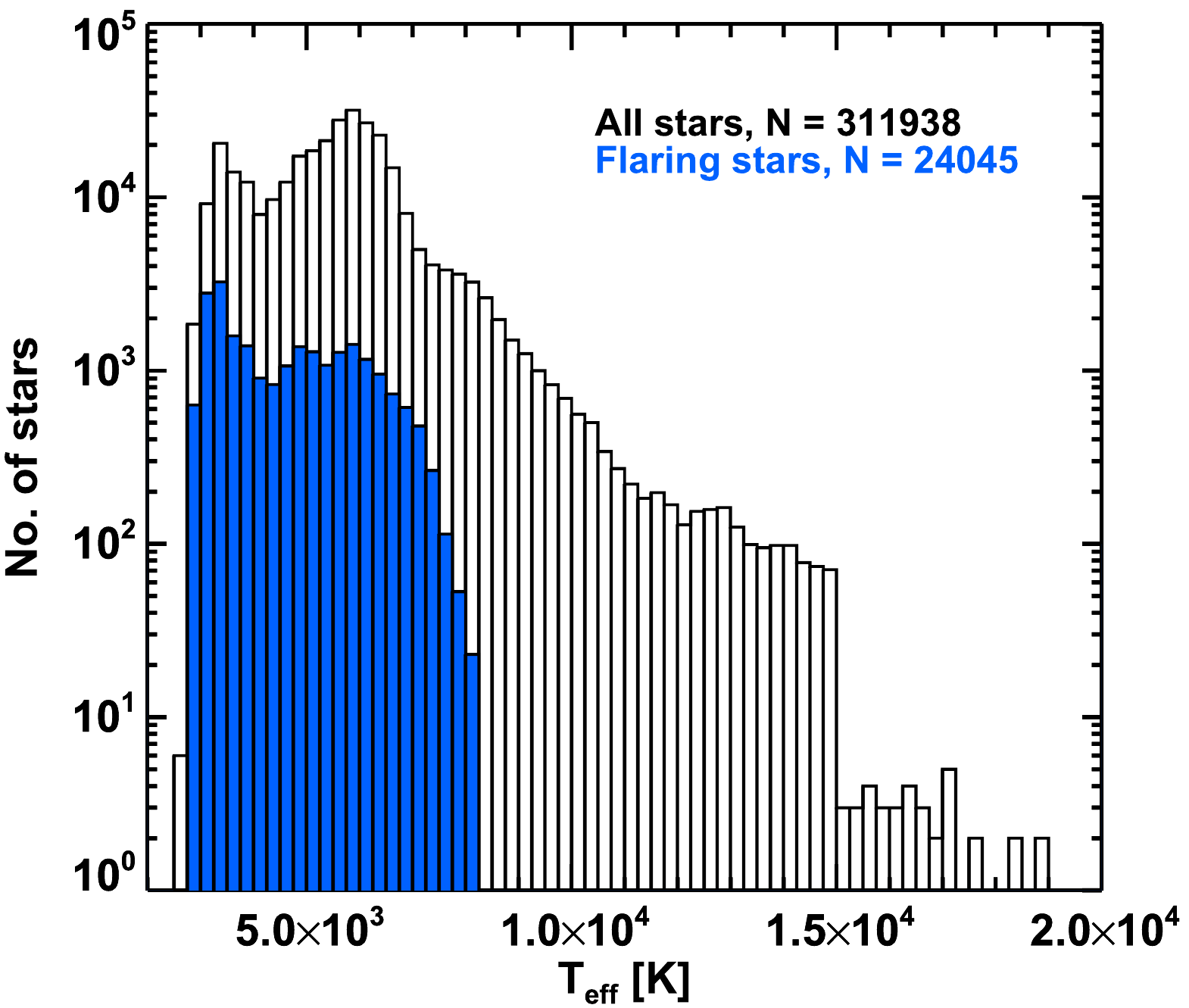}{0.348\textwidth}{}
          \fig{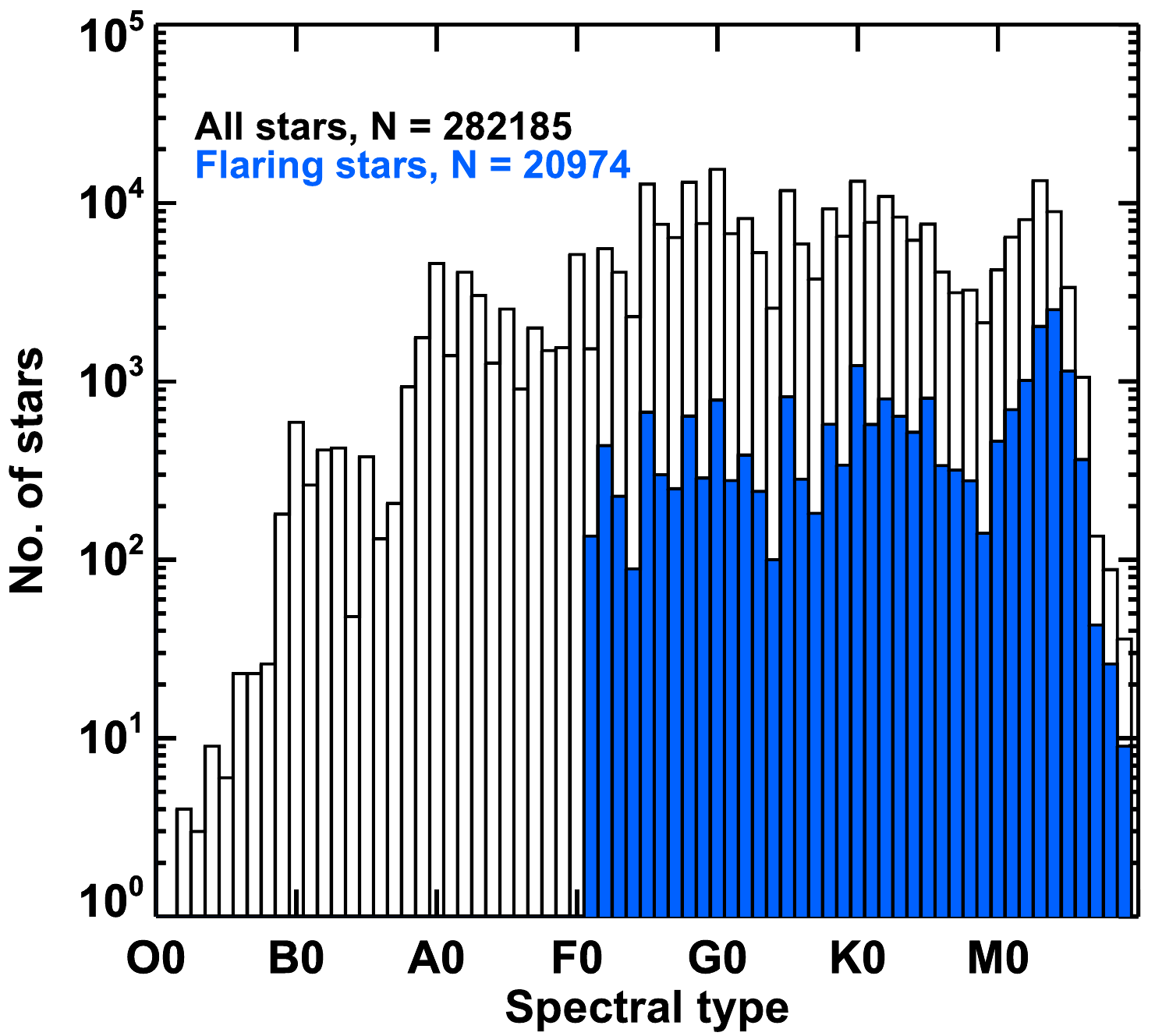}{0.33\textwidth}{}
          \fig{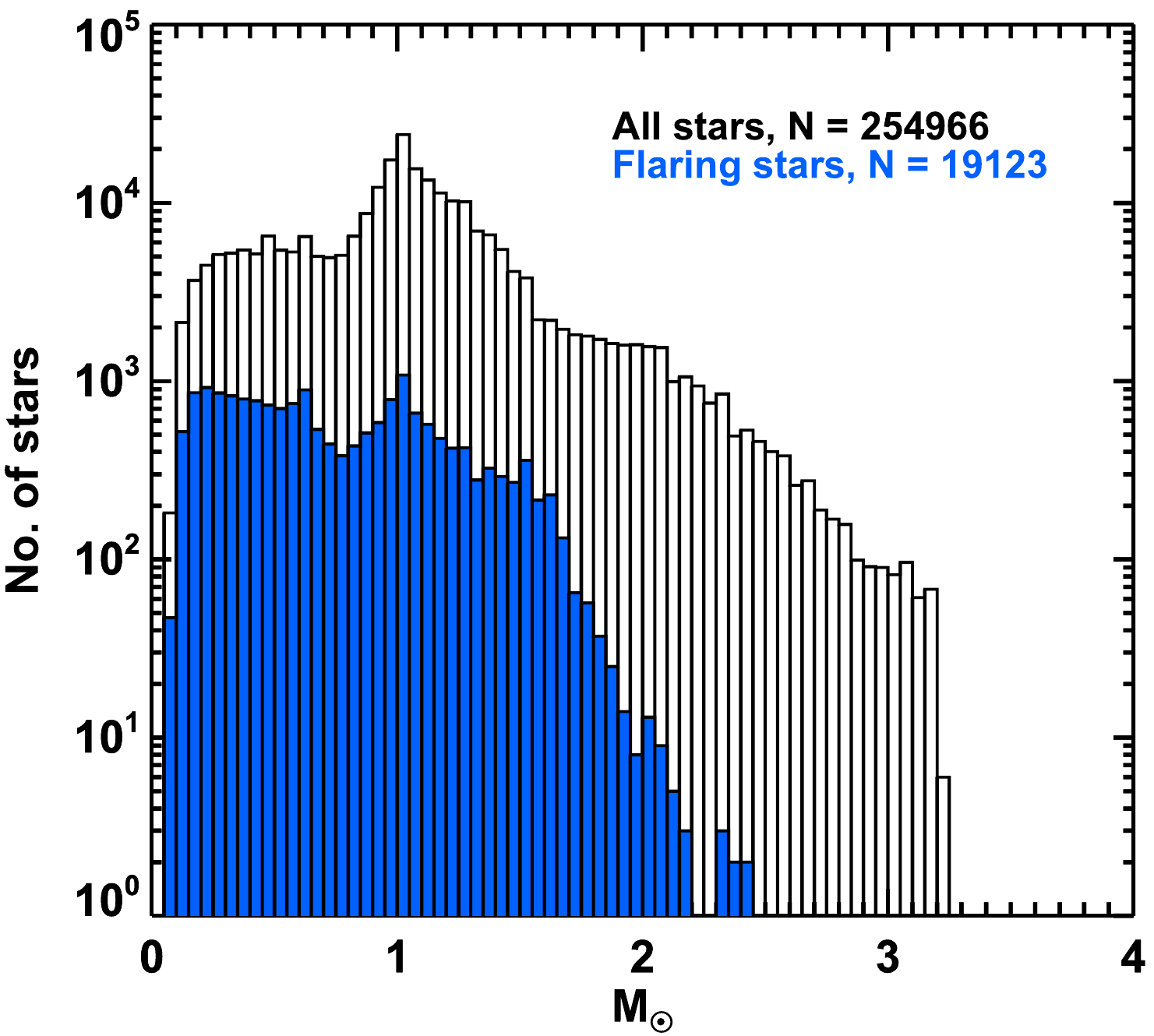}{0.33\textwidth}{}}
\caption{Histograms of the number of the flaring stars compared with the total number of stars in the \textit{TESS} observations of sectors 01-39 shown as a function of the stellar effective temperature (\emph{the left panel}), the stellar spectral type (\emph{the middle panel}) (in our work only flaring stars later than the F0 spectral type are analyzed) and the stellar mass (\emph{the right panel}).}
\label{fig:teff_mass_spt}
\end{figure}

\begin{figure}[H]
\gridline{\fig{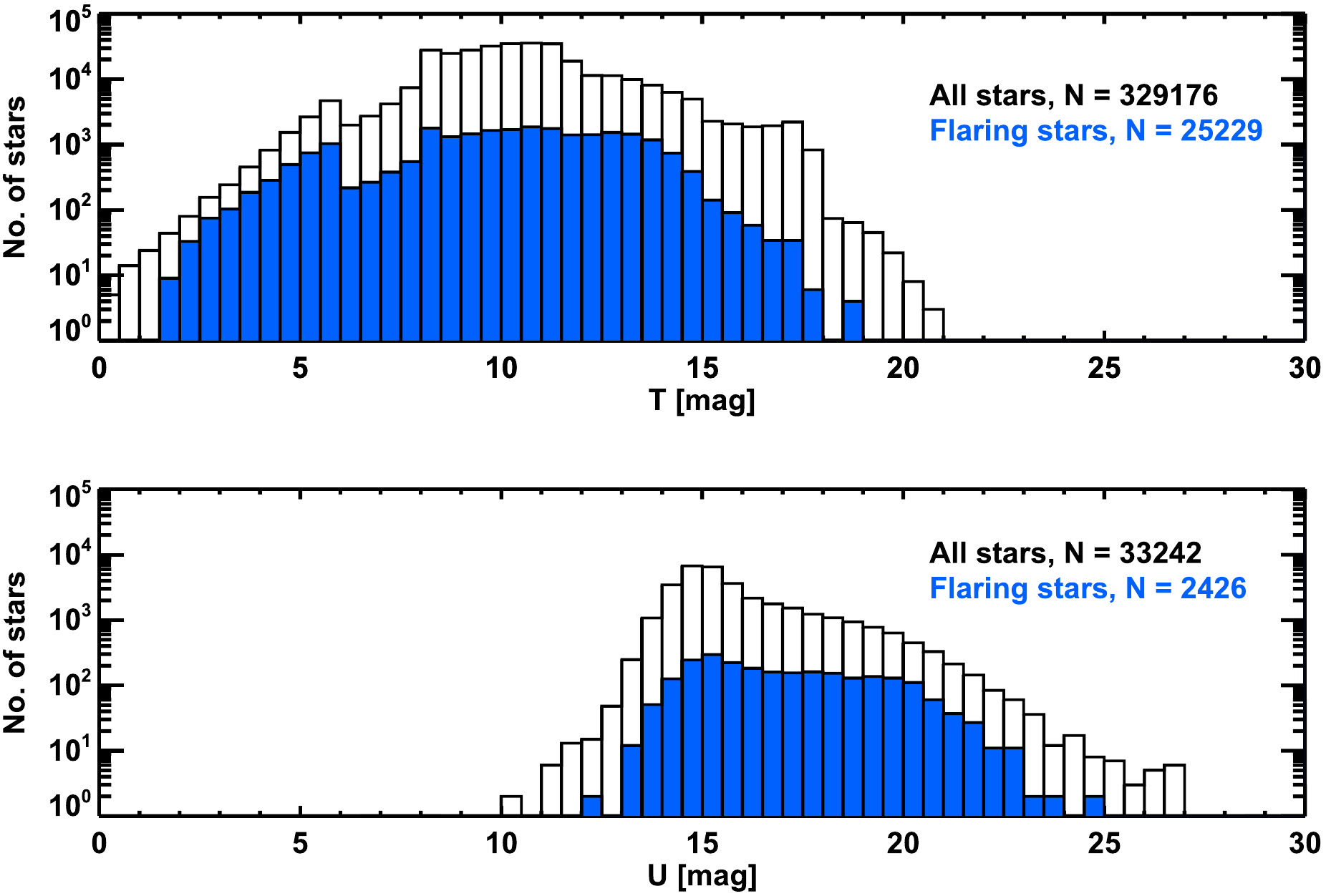}{0.49\textwidth}{}
          \fig{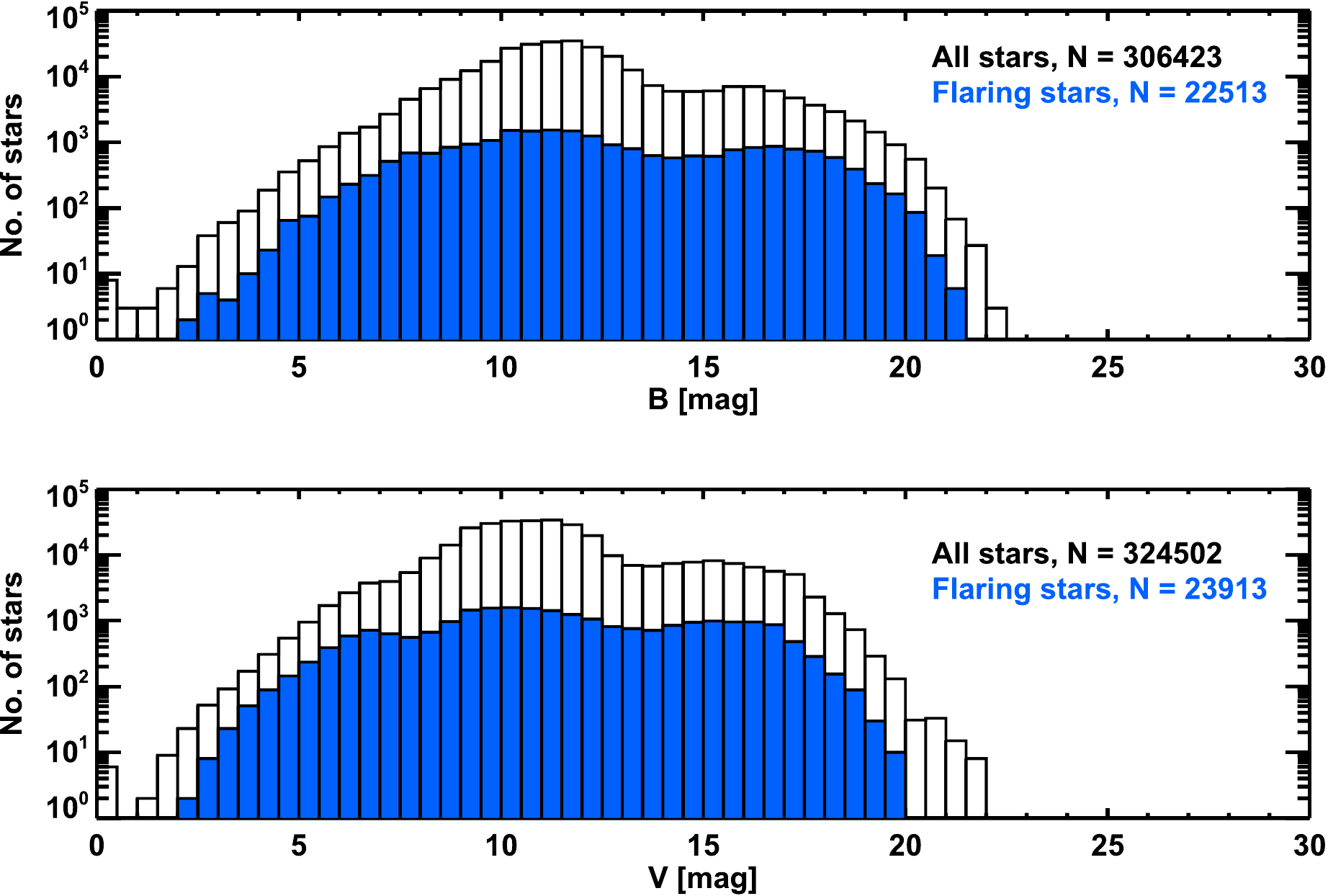}{0.49\textwidth}{}}
\caption{Histograms of the number of the flaring stars compared with the total number of stars in the \textit{TESS} observations of sectors 01-39 shown as a function of \textit{TESS} (\emph{the upper left panel}), B (\emph{the upper right panel}), U (\emph{the lower left panel}), V (\emph{the lower right panel}) magnitude.}
\label{fig:mag}
\end{figure}

The analysis carried out in this work allowed us to determine the distribution of the basic parameters of flaring stars, such as effective temperature, spectral type, mass, and magnitude (Figure \ref{fig:teff_mass_spt} and Figure \ref{fig:mag}). All analyzed stars are marked in gray, and flaring stars are marked in blue. The analyzed data sample of flaring stars includes only stars with spectral types later than F0. For this reason, most of these stars' effective temperatures do not exceed 8,000 K and 1.7 solar masses. As can be seen in Figure \ref{fig:teff_mass_spt}, the observed distributions of the presented parameters for the flaring stars are generally different than those for the entire sample of stars.

There is some similarity in the temperature distribution of the flaring stars, where the main maximum is about 3,000~K and the second one about 6,000 K. They are characteristic for the entire studied population. The distributions of flaring stars' spectral types and masses are dominated by stars with the spectral type M and masses of one solar mass or less. The distributions of \textit{TESS}, B, V, and U magnitudes for all stars and the flaring stars seems to be similar (Figure \ref{fig:mag}). However, based on the two-sample Kolmogorov–Smirnov test we reject the hypothesis that these samples are of the same distribution. We used samples of \textit{TESS}, B, V, and U magnitudes for all stars and only for the flaring stars. The distribution of U magnitude of the flaring stars is the least similar to the entire sample distribution due to the fact that the sample's of stellar U magnitude is only about 10\% of all stars observed by \textit{TESS}.

The parameters of the stars and the results obtained by us were examined using Benford's law \citep{10.2307/984802}. This was done to check whether the data and the results are uniformly and randomly distributed. The tests show that the catalog data are not similar to the Benford distribution. The results obtained by us, such as the duration of the flares' or estimated flares energy, are consistent with Benford's law.

\begin{figure}[H]
\centering
\includegraphics[width=0.65\textwidth]{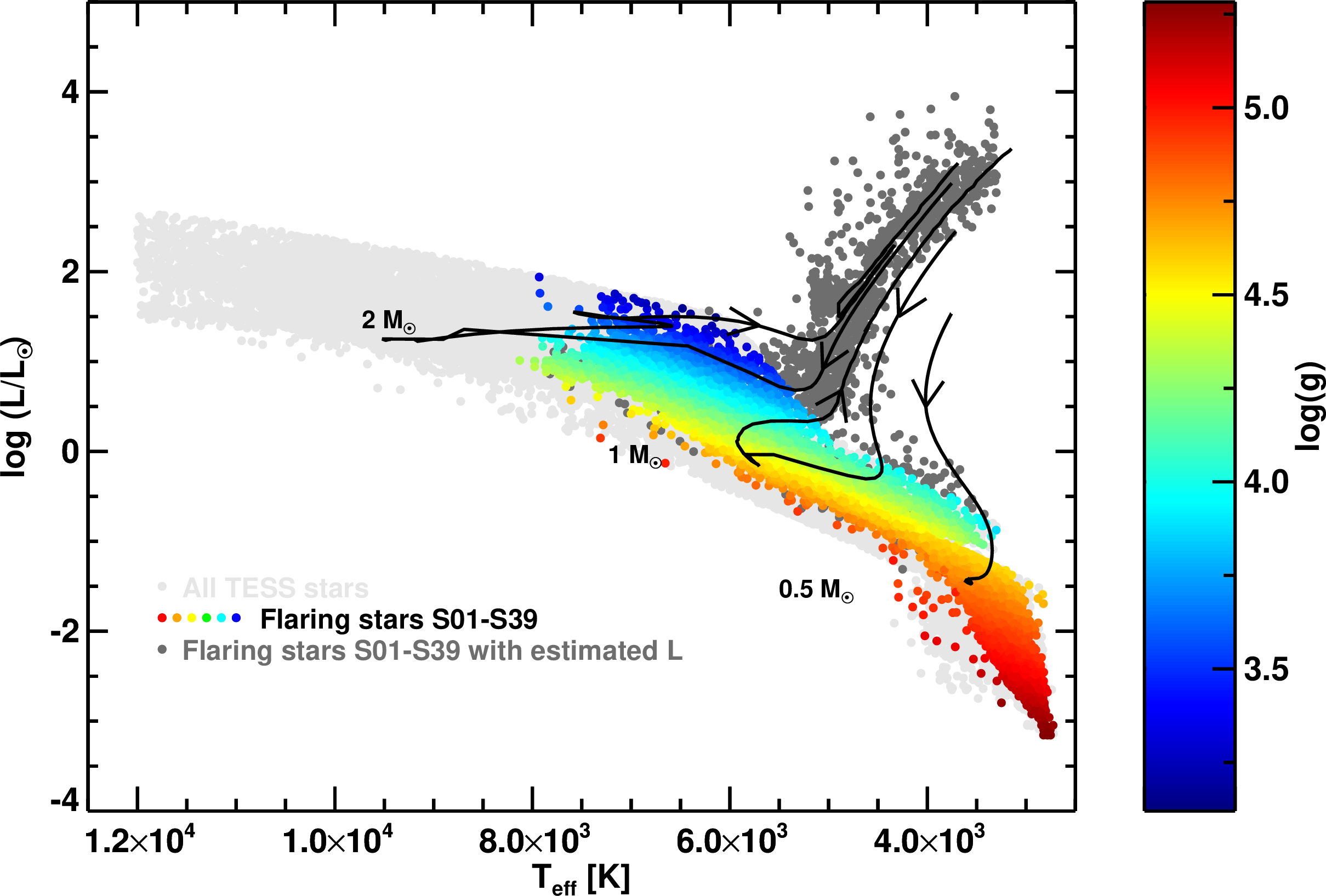}
\caption{H-R diagram for all stars observed by \textit{TESS} (light gray dots) and the flaring stars from sectors 01-39 with known luminosities and effective temperatures (colored dots). The flaring stars with estimated  luminosities are marked in dark gray.  On the right, there is the colorbar of stellar $log(g)$.  The black lines show the evolutionary tracks of stars with initial masses of 0.5, 1 and 2 masses of the sun.}
\label{fig:hr}
\end{figure}

All stars observed by \textit{TESS} with given luminosity and effective temperature were marked as light gray dots on the Hertzsprung-Russell (H-R) diagram (Figure \ref{fig:hr}). The flaring stars with known luminosities, effective temperatures, and $log(g)$ observed in sectors 01-39 are marked with colors. The flaring stars with estimated luminosity and no $log(g)$ are marked in dark gray. The luminosity of the star was estimated from the stellar radius and the effective temperature. On the right, there is the colorbar of stellar $log(g)$. The $log(g)$ values of the flaring stars range from 3.3 to 5.1. Solid black lines represent the evolutionary tracks of stars with an initial mass of  0.5, 1, and 2 solar masses. Most of the flaring stars that we found are in the main-sequence, but we detected flares on young stars and giants as well. To estimate the age of the star we use catalog data from SIMBAD. We consider an object as a young if the star is classified as a pre main-sequence, a young stellar object, T Tauri, or Wolf-Rayet star. The energies of the flares on young stars range from $8 \times 10^{31}$ to $4 \times 10^{35}$ erg. We compared the estimations of the energy of several super-flares on young stars with that in \citet{2020Feinstein}. For example, we determined the flare energy on TIC44678751 to be $2 \times 10^{35}$ erg and \citet{2020Feinstein} to be $2 \times 10^{37}$ erg. There are similar big differences for other events, but we are not able to clearly state what they may result from. The two orders of magnitude difference cannot be explained only by assumed flare temperatures (10,000 K in our case and 9,000 K in \citet{2020Feinstein}). The temperatures of the flaring stars mostly do not exceed 8,000 K, but we found a few stars with higher effective temperatures.

\begin{figure}[h]
\gridline{
\fig{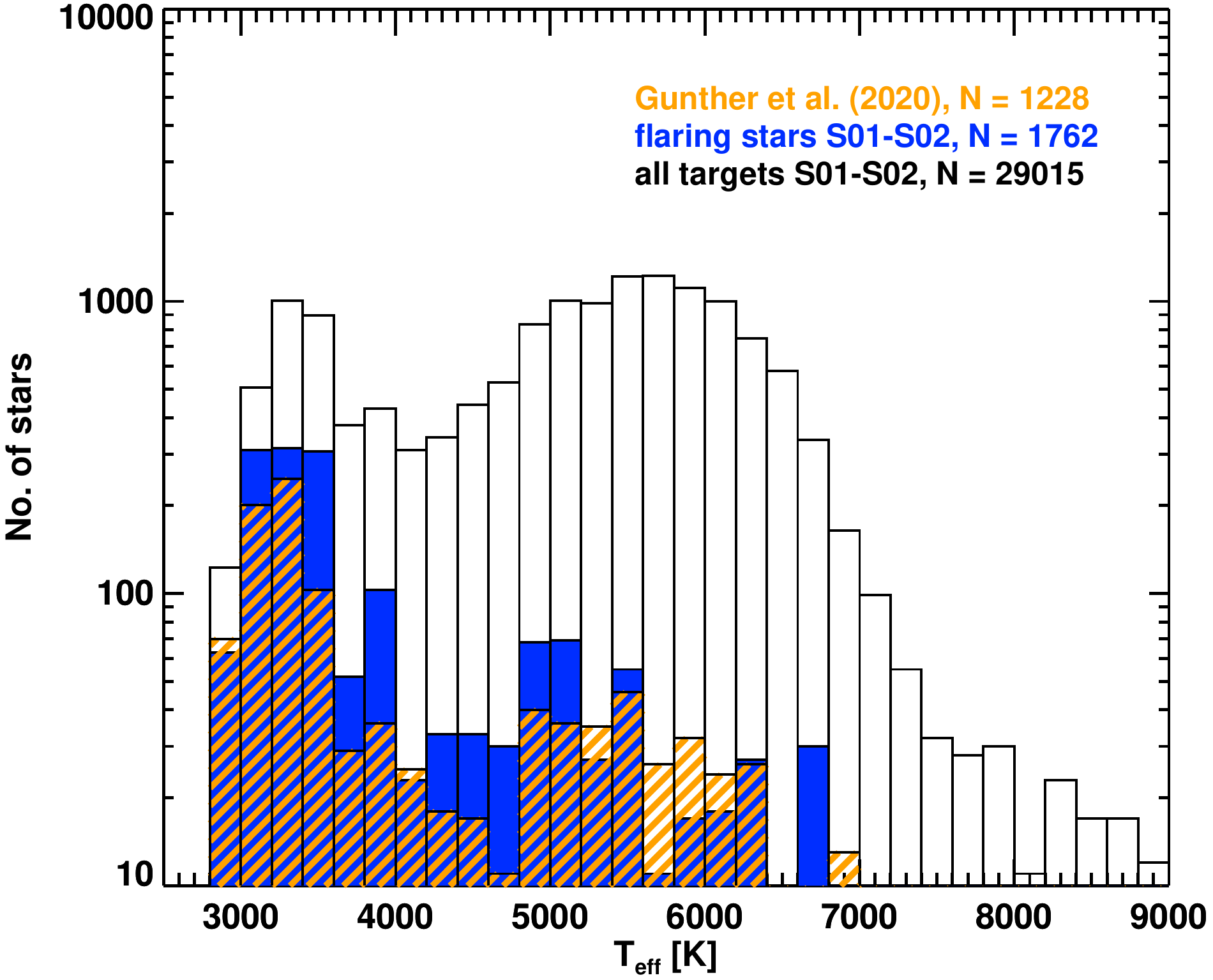}{0.49\textwidth}
{\caption{Histogram of the number of flaring stars compared with the total number of stars in the \textit{TESS} two-min cadence observations of sectors 1 and 2 (S01-S02) shown as a function of the stellar effective temperature. Comparison of our results and \citet{gunther}.} \label{fig:lit11}}
\fig{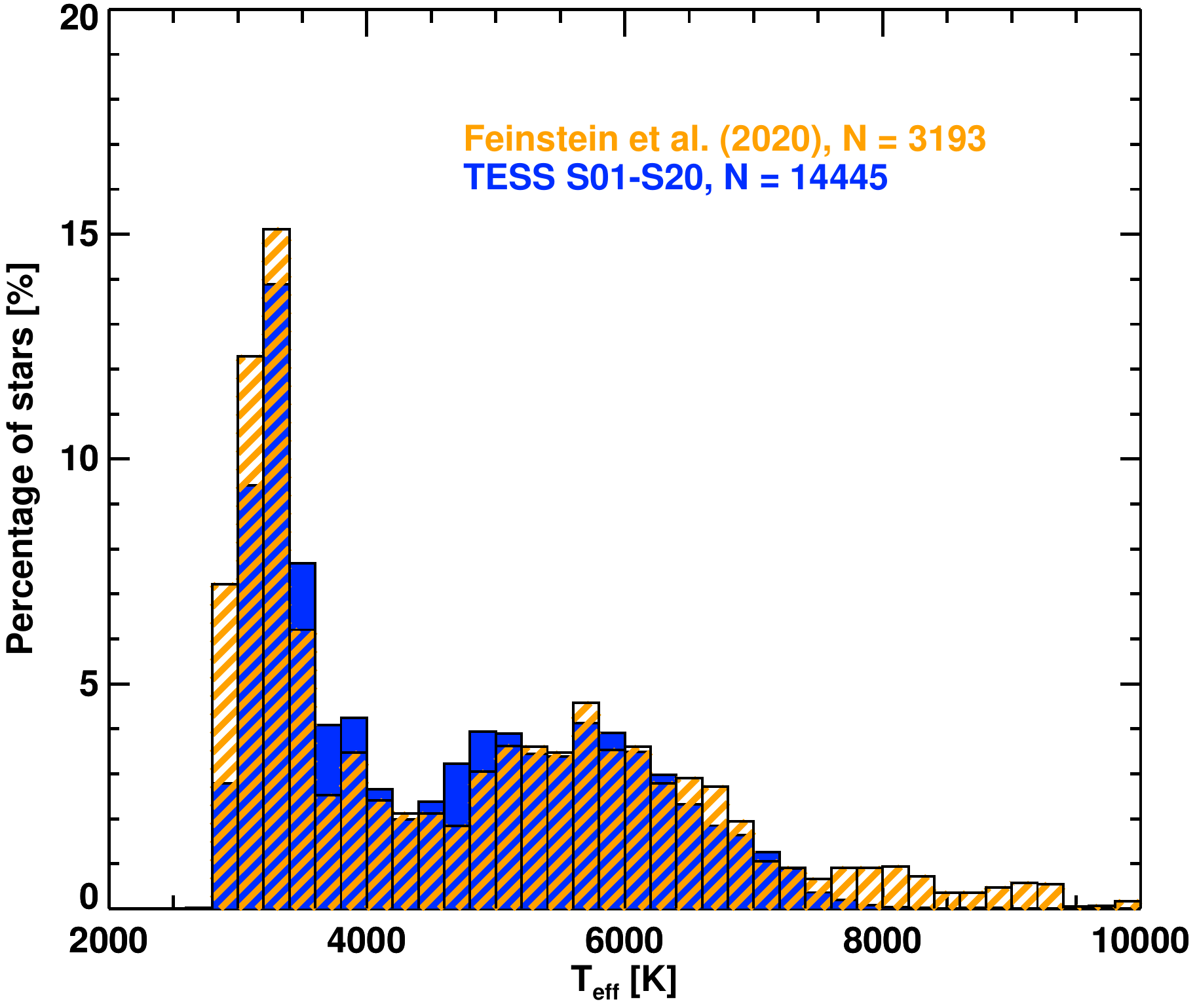}{0.472\textwidth}
{\caption{The distribution of $T_{\mathrm{eff}}$ within stars observed in two-min cadence by \textit{TESS} in sectors 1-20 (S01-S20). Comparison of our results and \citet{2020Feinstein}.} \label{fig:lit12}}}
\end{figure}

\begin{figure}[H]
\gridline{
\fig{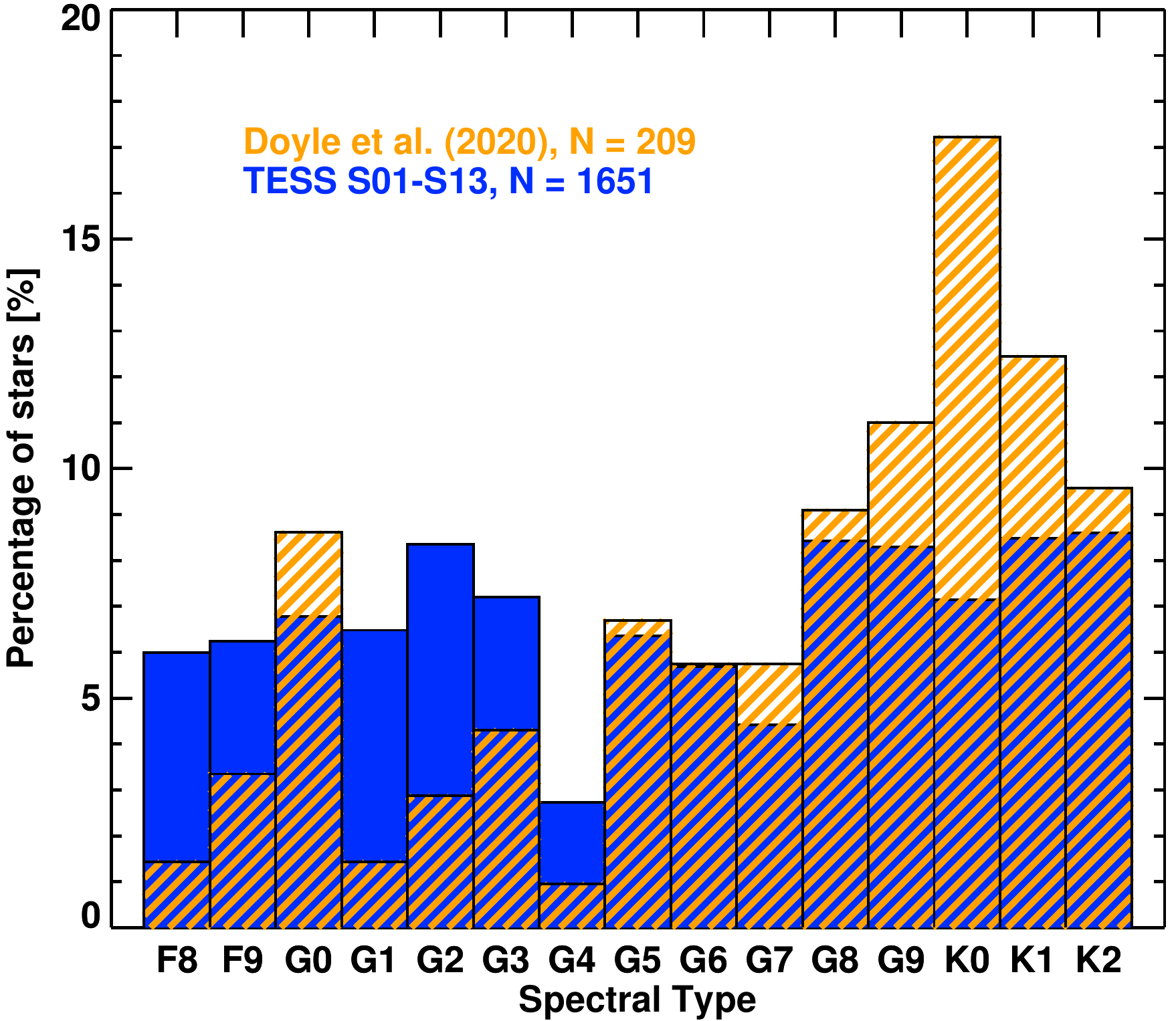}{0.471\textwidth}
{\caption{A histogram showing the spread in spectral types within the solar-type star sample observed in two-min cadence by \textit{TESS} in sectors 1-13 (S01-S13). Comparison of our results and \citet{2020MNRAS.494.3596D}.} \label{fig:lit21}}
\fig{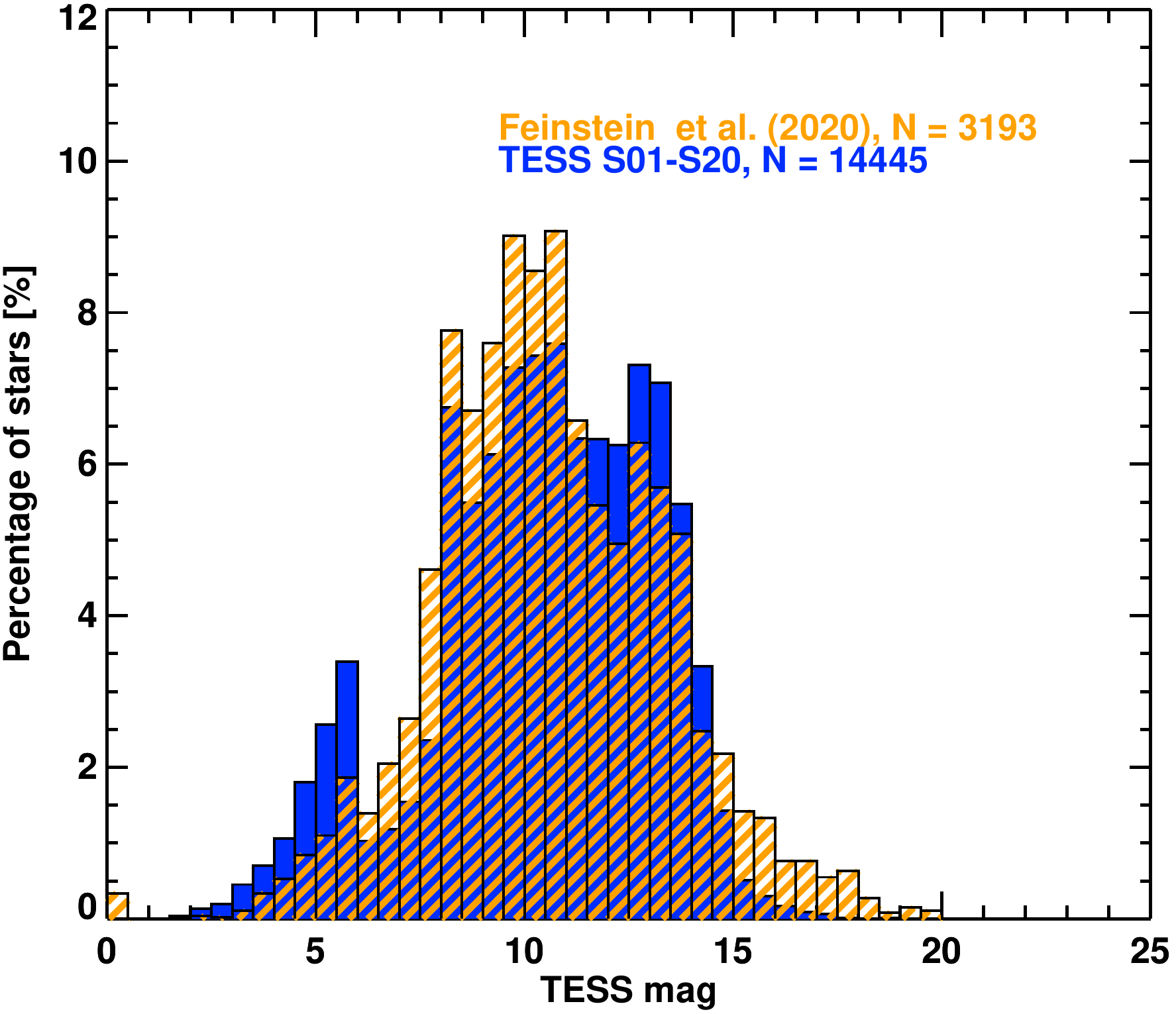}{0.4755\textwidth}
{\caption{The distribution of \textit{TESS} magnitudes within the star sample observed in two-min cadence by \textit{TESS} in sectors 1-20 (S01-S20). Comparison of our results and \citet{2020Feinstein}.} \label{fig:lit22}}}
\end{figure}

The results from our work (marked in blue in Figure \ref{fig:lit11}, \ref{fig:lit12}, \ref{fig:lit21}, \ref{fig:lit22}) were compared with works such as \citet{2020MNRAS.494.3596D}, \citet{2020Feinstein}, \citet{gunther}. However, there are some differences in the sample selection between that works and our research. In \citet{2020MNRAS.494.3596D} authors researched a solar-type stars with measured rotation periods observed in TESS sectors 1-13. \citet{gunther} was analyzing flares at all 2-minute cadence stars only in TESS sectors 1 and 2. \citet{2020Feinstein} was looking at only young stars in sectors 1-20. Figures \ref{fig:lit11} and \ref{fig:lit12} show histograms of the number of flaring stars compared with results from \citet{gunther} and \citet{2020Feinstein}  presented as a function of the stellar effective temperature. To make the comparison, we selected all the stars from  \citet{gunther} form Table 2, that have estimated the effective temperature. The software prepared by us found more flaring stars in sectors 1 and 2 than in \citet{gunther}, mainly due to different methods of searching for flares. In both samples there are 963 flaring stars with the same TIC number and 7837 flares were found by both pipelines. However, it should be noted that our software analyzes only events that have at least six observation points. The effective temperature of flaring stars distribution fits nicely. Using the  ${\chi}^2$ goodness-of-fit test between two distributions we obtain the value of about 0.90. There is no reason to reject the hypothesis at the 0.05 significance level. Compared to the \citet{2020Feinstein} (Figure \ref{fig:lit12}), the temperature distributions of flaring stars are similar up to 8,000 K (${\chi}^2$ goodness-of-fit test value is about 0.93). We also chose the same sample that was presented by the authors in their work. The effective temperatures of most flaring stars that we have found are limited to this value due to the assumptions described earlier. There is a large difference in both samples' sizes, therefore on the vertical axis is a relative number of stars. 

Figure \ref{fig:lit21} shows a spread in spectral types within solar-type star samples from our work and \citet{2020MNRAS.494.3596D} (observations from sectors 1-13). Again, the relative number of stars can be found on the vertical axis. The difference in numbers of stars is due to the sample selected in \citet{2020MNRAS.494.3596D}, where the spot modulation was searched for. The different distribution of the spectral types of flaring stars may result from the different methods of searching for flares used. In \citet{2020MNRAS.494.3596D} the sample of the studied stars was small enough to check visually each light curve, while our detections are mainly from using the automated software WARPFINDER. 

In Figure \ref{fig:lit22} we compare the distributions of \textit{TESS} magnitudes within the flaring stars observed in sectors 1-20 from our work and \citet{2020Feinstein}. As in Figure \ref{fig:lit12}, slight differences in the \textit{TESS} magnitude distribution are probably caused by the different methods of detecting flares used and limitations of the stellar age up to 800 Myr.

\subsection{Parameters of stellar flares}

\begin{figure}[H]
\gridline{\fig{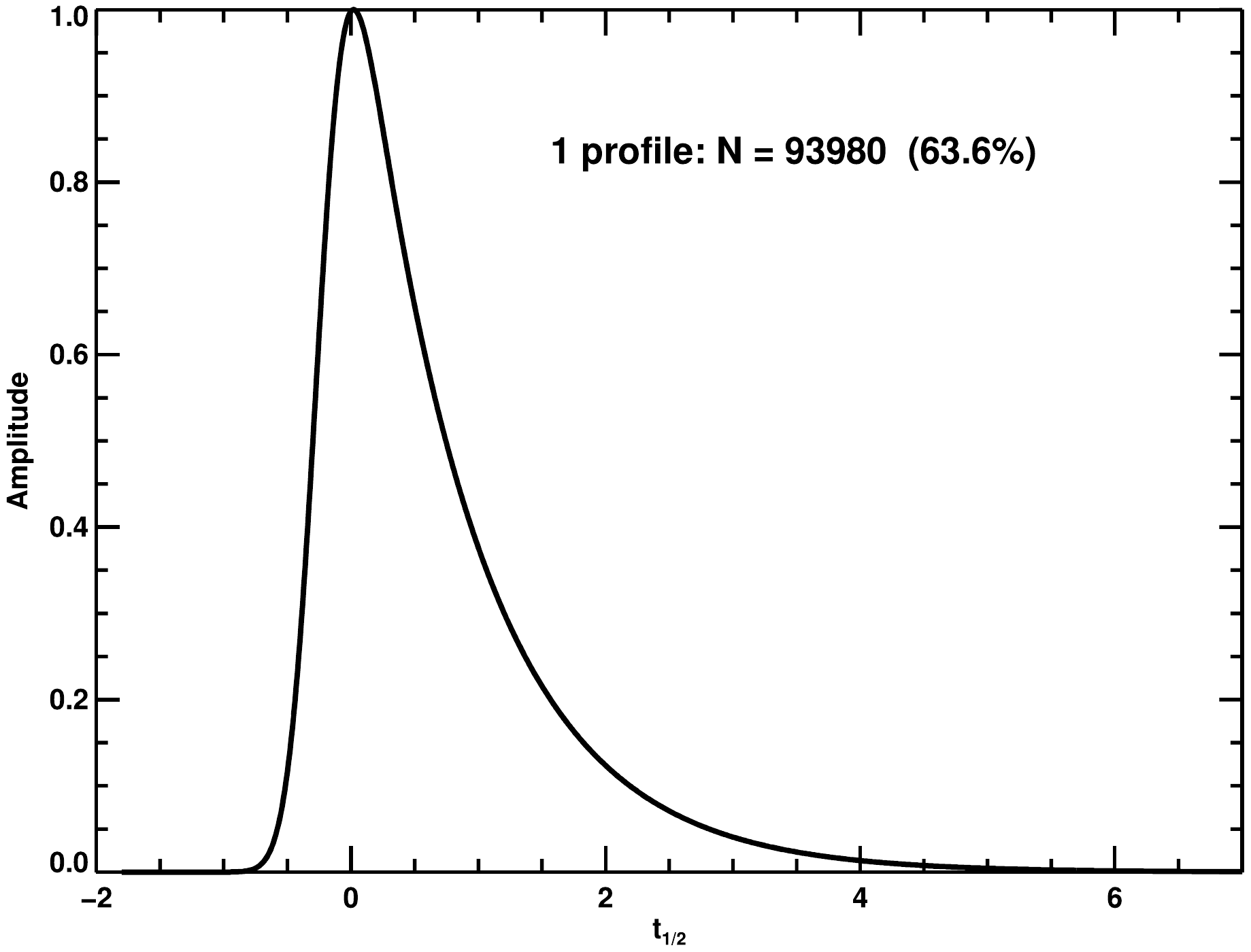}{0.47\textwidth}{}
          \fig{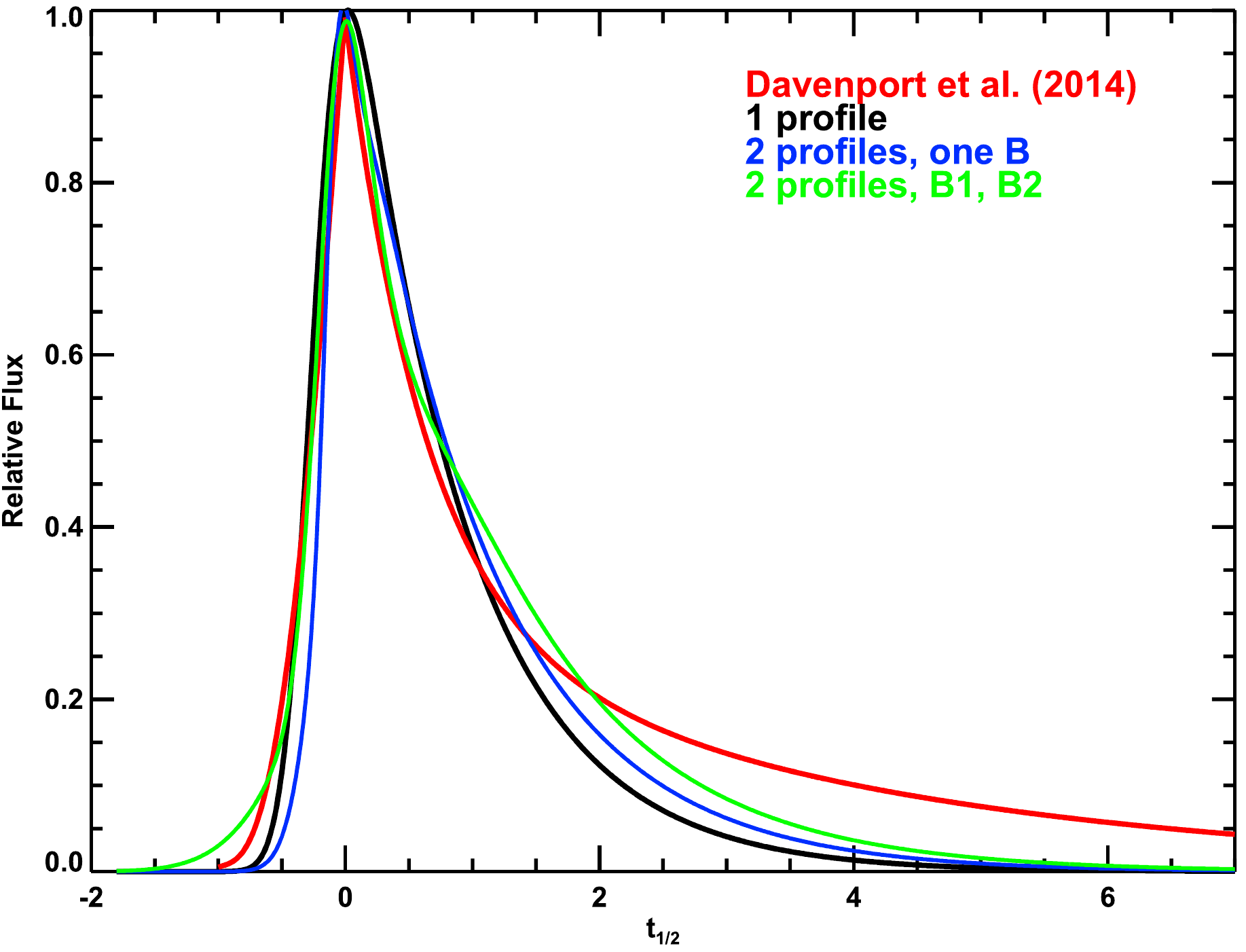}{0.47\textwidth}{}}
\caption{\emph{Left:} Average flare profile described by one profile. \emph{Right:} The profiles obtained in our study (one profile - black, two profiles with one B - blue, two profiles with B1 and B2 - green) compared with the profile from \citet{Davenport_2014} (red).}
\label{fig:flares_prof}
\end{figure}

\begin{figure}[H]
\gridline{\fig{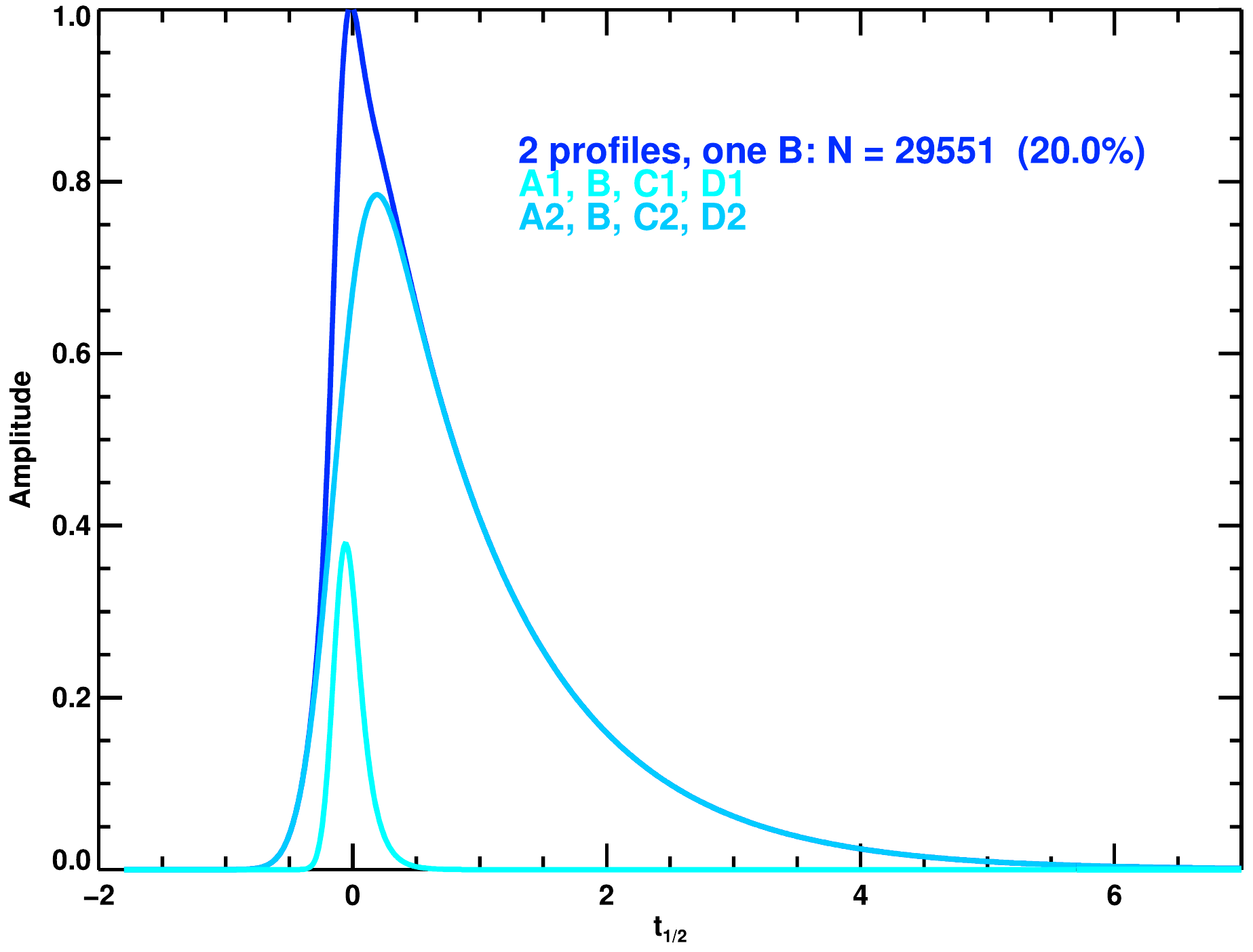}{0.47\textwidth}{}
          \fig{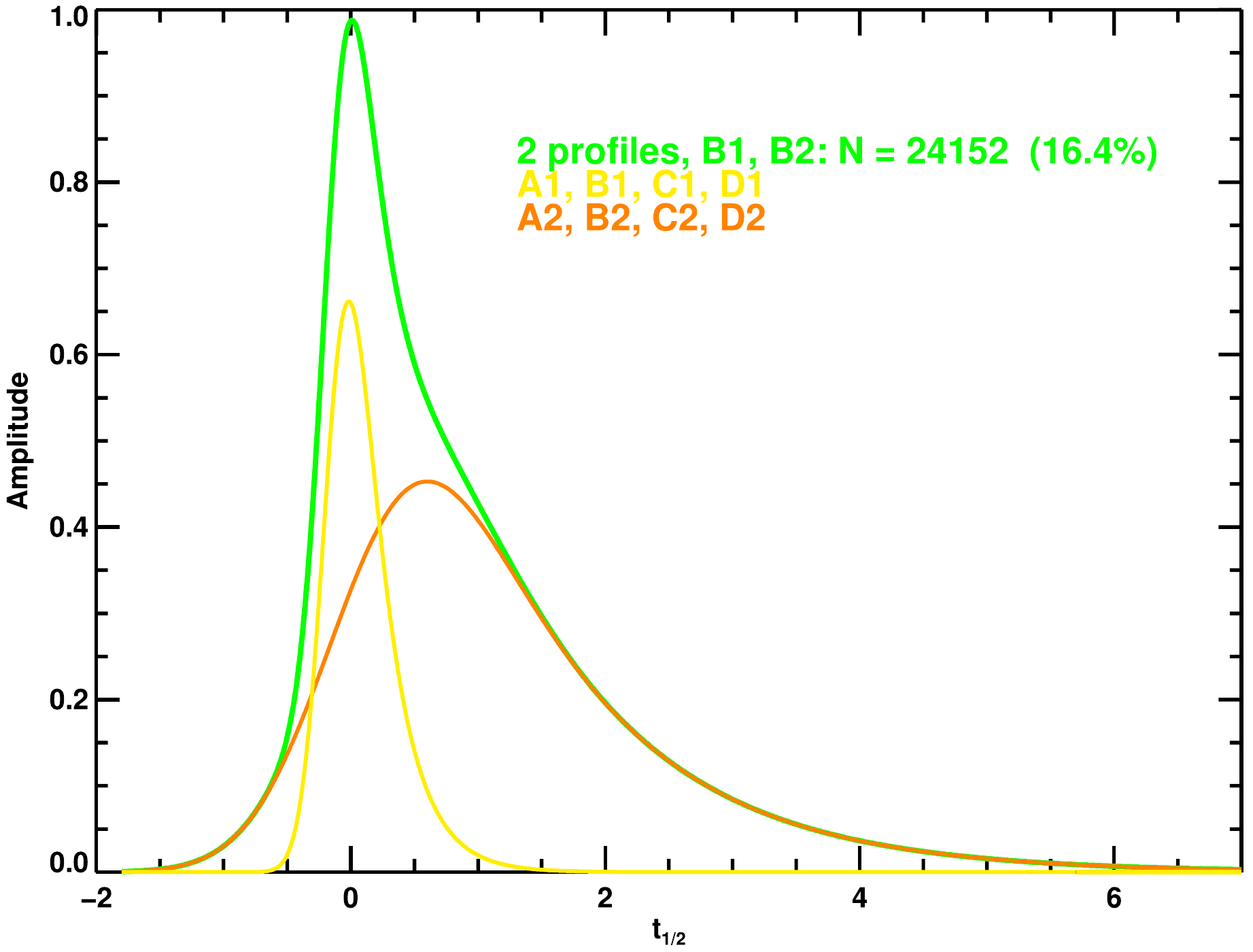}{0.47\textwidth}{}}
\caption{\emph{Left:} Average flare profile with one B (dark blue) described by two components (light blue). \emph{Right:} Average flare profile with B1 and B2 (green) described by two components (yellow and orange).}
\label{fig:flares_prof2}
\end{figure}

\begin{figure}[H]
\centering
\includegraphics[width=0.52\textwidth]{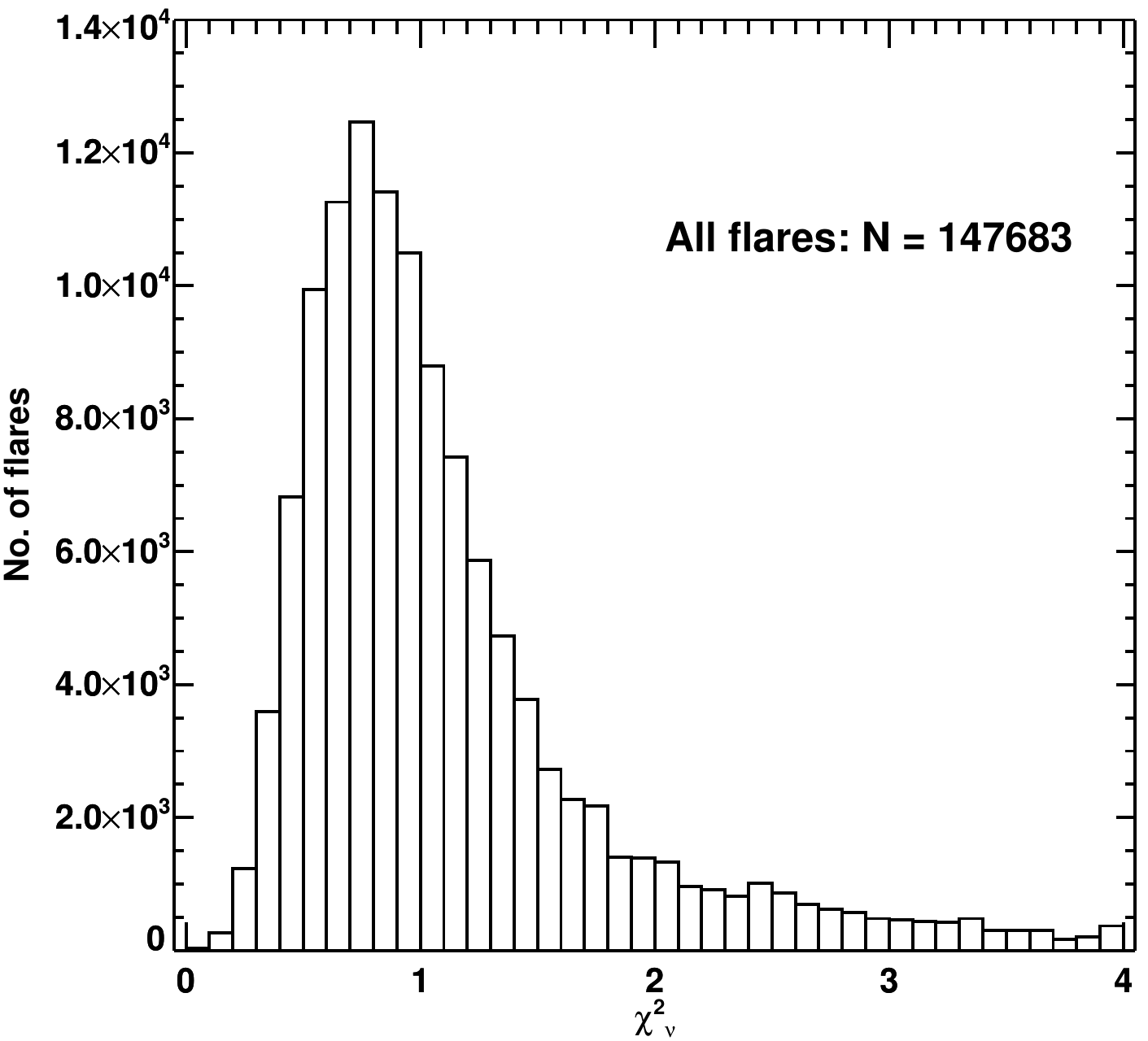}
\caption{The distribution of the values of the reduced ${\chi}^2$ for the analyzed stellar flares.}
\label{fig:chi}
\end{figure}

Light curves of the detected flares were fitted with one of three types of profiles presented in the description of the flare profile method. We used the single profile, double with the same value of B and double with the different value of B. More than half of the events fitted well with only the single profile. In other cases two profiles were needed for a good fit of a single flare. In this analysis, we did not handle the problem of fitting two or more overlapping events. The overlapping events do not fit well by the profile we are using (Equations \ref{eq:profile1}, \ref{eq:profile2_1B}, \ref{eq:profile2_2B}), so the ${\chi}^2$ value is too high. For this reason, they are automatically rejected from the analysis. Based on the well-fitted flares (the value of the reduced ${\chi}^2$ less than 3.0), we reconstructed three types of average profiles as a function of full-time width at half the maximum flux ($t_{1/2}$) (Figure \ref{fig:flares_prof}, Figure \ref{fig:flares_prof2}). All flares were previously scaled to a relative time and amplitude. We then fitted the  parameters A, B, C, and D of the obtained average profiles (Table \ref{tab:table_abcd}). A similar method was used in \cite{2013Kowalski} and \citet{Davenport_2014}. The distribution of the values of the reduced ${\chi}^2$ for the analyzed stellar flares is shown in Figure \ref{fig:chi}.

\begin{deluxetable}{ccccc}[H]
\tablenum{2}
\tablecaption{The parameters A, B, C, and D of average flares profiles
\label{tab:table_abcd}}
\tablewidth{0pt}
\tablehead{
\colhead{} & \colhead{A} &  \colhead{B} &
\colhead{C} & \colhead{D}}
\startdata
1 profile     & 3.02   &    1.55   &   0.26    &   1.11              \\ 
2 profiles, one B (1)     & 5.08    &  1.68  &   0.11   &   9.65  \\ 
2 profiles, one B (2)    &   2.15  &   1.68 &     0.30 &    0.94        \\
2 profiles, B1, B2 (1)    & 3.96    &   1.64    &  0.22   &  3.94       \\ 
2 profiles, B1, B2 (2)  & 0.65   &    1.81 &   0.81  &    0.84            \\ 
\enddata
\end{deluxetable}

The profiles differ in the average growth time (parameters B and C) and the decay time (parameter D). Longer and stronger events are usually fitted with two profiles. One of these profiles is supposed to be related to the direct heating of the photosphere by non-thermal electrons. It is characterized by very short growth and decay times (parameters B, C and D). The second one, longer (with significantly lower parameter D) may be the result of back-warming processes \citep{2007Isobe}. It can be noticed that the contribution of the second component is particularly important for a profile with one B parameter. In both double profiles, the second component dominates in the decay phase.

The flare profiles that we obtained were compared with the flare model form \citet{Davenport_2014}, which is the result of the analysis of flares observed with \textit{Kepler} on the M4 star GJ 1243. Figure \ref{fig:flares_prof} (right) shows our three profiles marked with black, blue and green, and the reconstructed model of a flare proposed by \citet{Davenport_2014} (red). This analytic model was forced to go through 0 at $t_{1/2}=-1$ and 1 at $t_{1/2}=0$. The large differences between the profiles at the decay phase ($t_{1/2} > 2$) may be related to different methods of determining the end times of stellar flares.

\begin{figure}[H]
\gridline{\fig{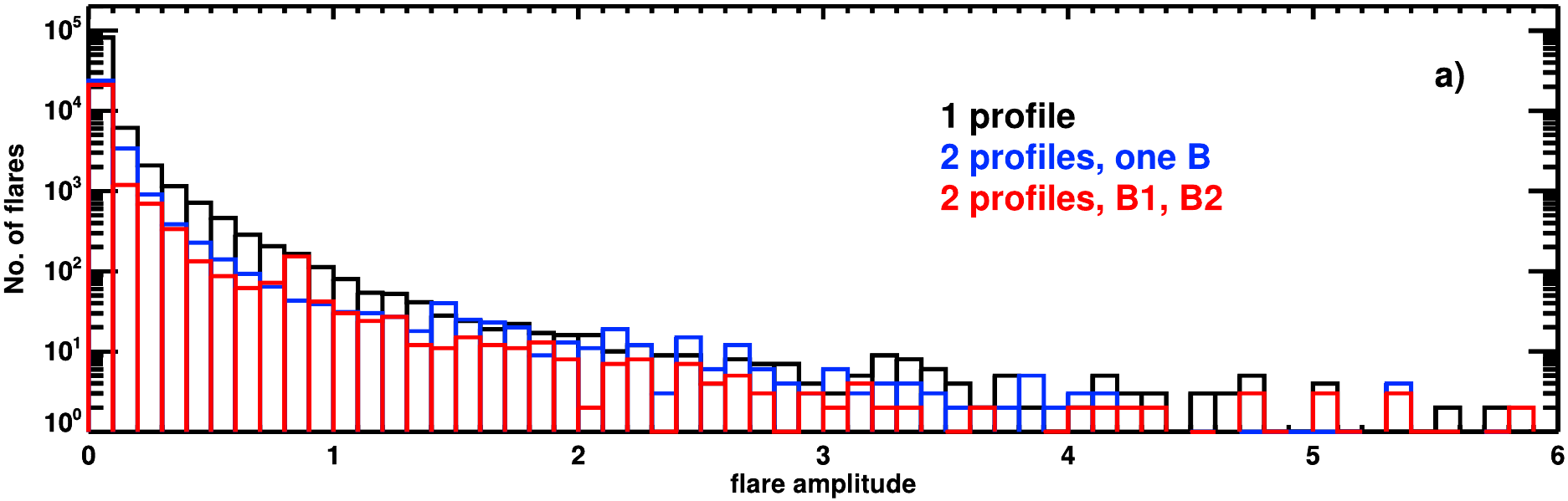}{0.48\textwidth}{}
          \fig{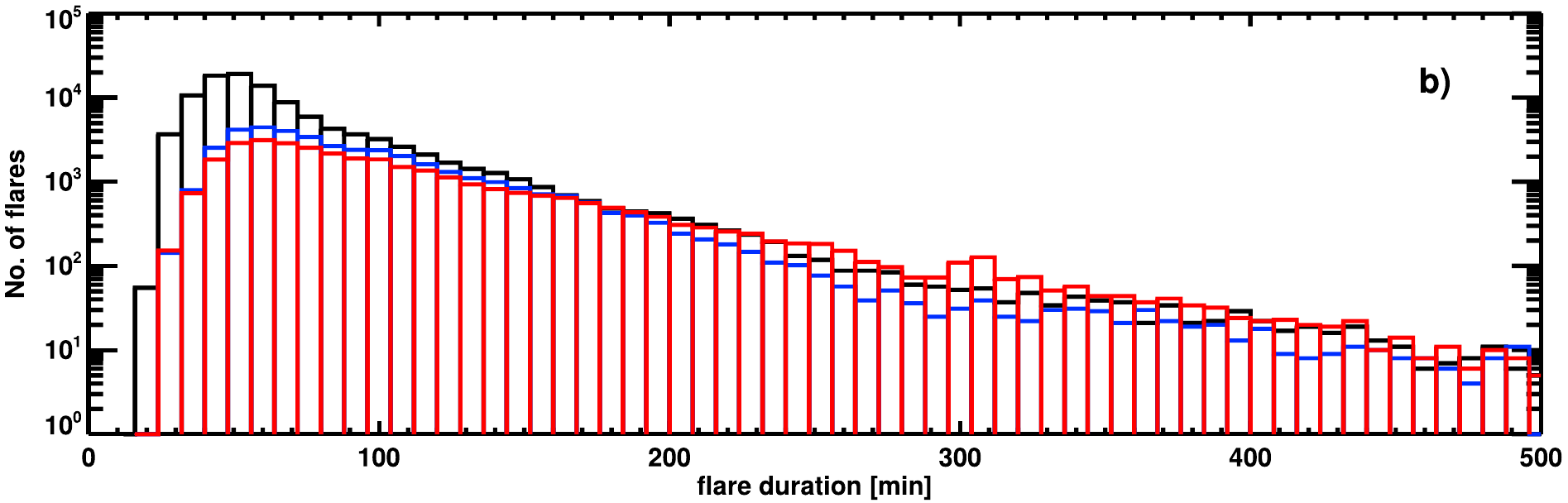}{0.49\textwidth}{}}
\gridline{
          \fig{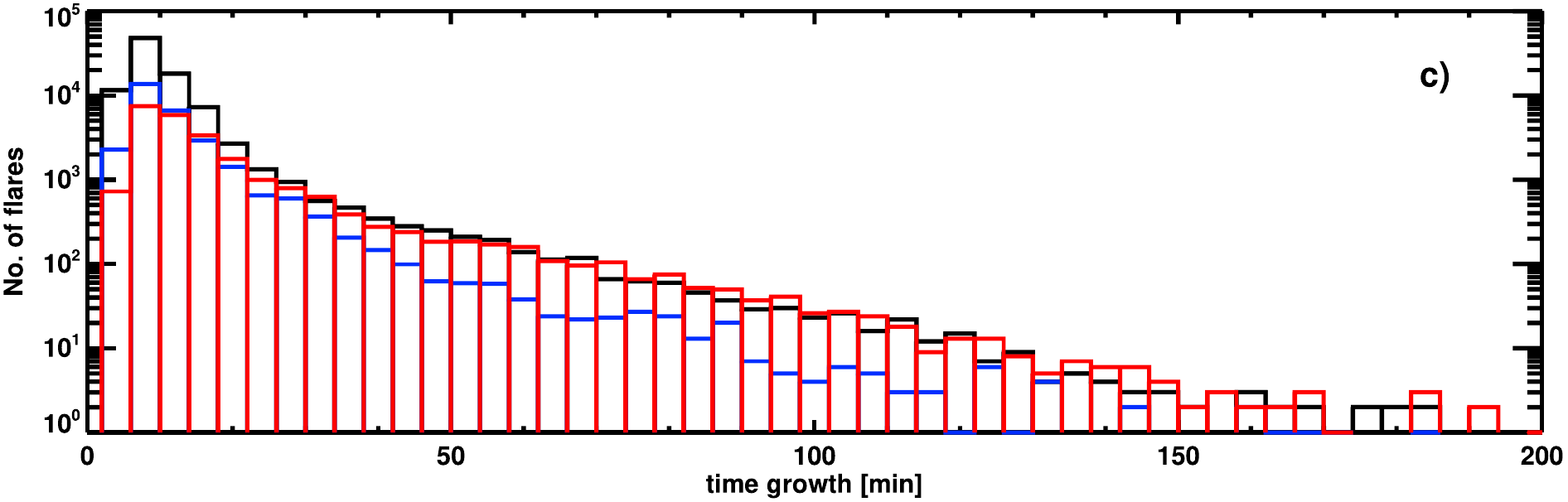}{0.49\textwidth}{}
          \fig{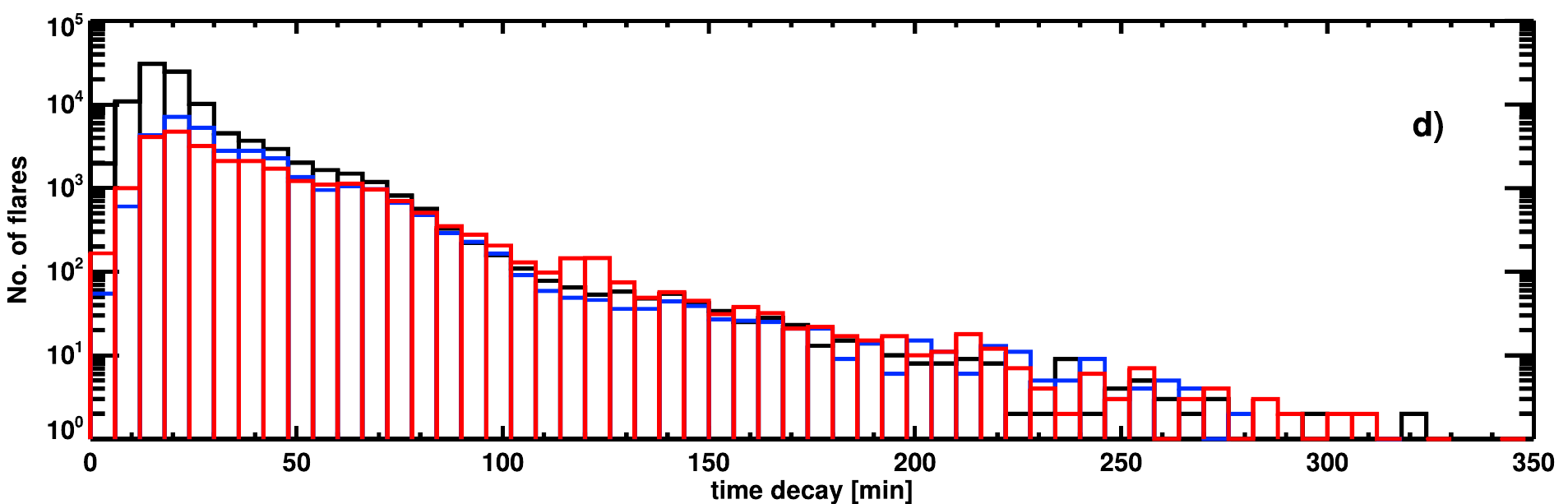}{0.49\textwidth}{}}
\caption{Distributions of the basic parameters of stellar flares: a) amplitude, b) duration, c) growth time, and d) decay time. The flares fitted with one profile are marked in black, with two profiles with one B - in blue, and with two profiles with B1 and B2 - in red.}
\label{fig:flares_param}
\end{figure}

We examined the distribution of the basic parameters of stellar flares. Figure \ref{fig:flares_param} shows the distributions of flare amplitudes, durations, and times of growth and decay. The amplitudes of most of the detected events do not exceed the value of 0.1 of the normalized flux, however, there are flares with amplitudes equal even several times the brightness of the star. The maximum of the all flares' durations distribution is approximately 50 minutes. The flares fitted with two profiles are characterized by a longer average duration. In this work, we limited the minimum of a flare duration to six data points (12 minutes). The longest observed events last up to several hours. Most of the detections with a duration of more than 500 minutes were false, so we did not include them in further analysis. The average values of the growth times are much shorter than the decay times and are usually below 20 minutes. Examples of flares of long durations, times of growth or decay are listed in Table \ref{tab:table_times}.

\begin{deluxetable}{ccccccc}[H]
\tablenum{3}
\tablecaption{Table with examples of flares of the long duration or decay times
\label{tab:table_times}}
\tablewidth{0pt}
\tablehead{
\colhead{ID} & \colhead{Name} &  \colhead{T$_{\rm{eff}}$} &  \colhead{Duration} &  \colhead{Growth time}  & \colhead{Decay time} & \colhead{Amplitude}\\
\colhead{} & \colhead{} & \colhead{[K]} & \colhead{[min]}  & \colhead{[min]} & \colhead{[min]} & \colhead{[rel. flux]} } 
\startdata
TIC236778955 $^1$      & TYC 3926-1315-1 & 4279   & 350 & 90 & 240  &  0.356 \\ 
TIC236778955 $^1$      & TYC 3926-1315-1 & 4279   & 308 & 84 & 224  &  0.058 \\ 
TIC219218697 $^2$       & AG Dor  & 4851  & 256           & 30         & 226 & 0.011  \\ 
TIC152176559 $^2$     & V1217 Cen & 3639   & 234       & 62      &  172  &  0.048 \\
TIC149915491 $^1$     & RX J1039.7+6545 & 3704 & 222      & 42        & 180    & 0.030
\enddata
\tablenotetext{\tiny 1}{  Probably no overlapping events}
\tablenotetext{\tiny 2}{  Appear to be overlapping events}
\end{deluxetable}

\begin{figure}[H]
\gridline{
\fig{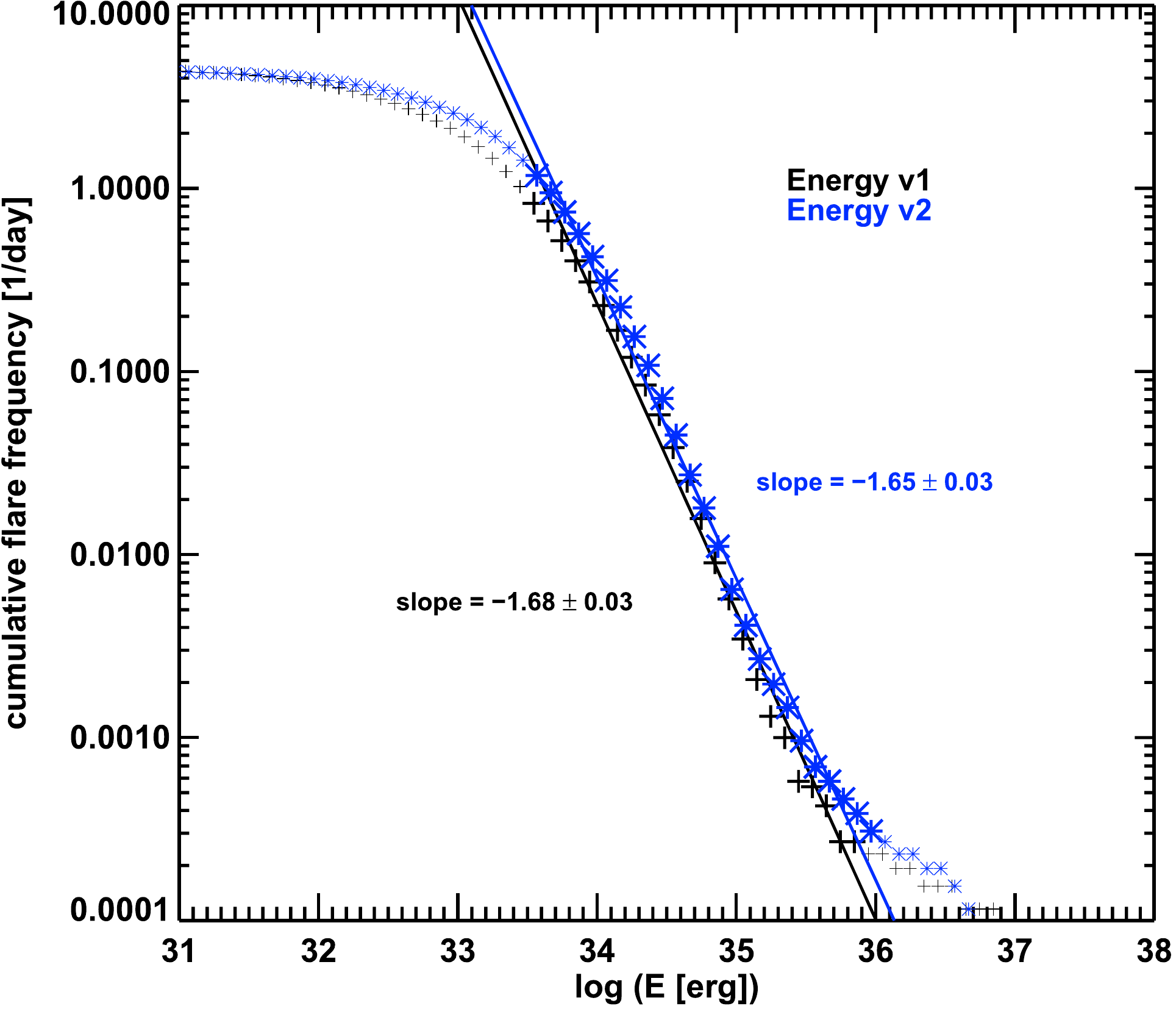}{0.4\textwidth}
{\caption{Cumulative distribution of the flare energies from \textit{TESS} sectors S01-S39. The energies, the fitted power function and the power-law index estimated using the method based on \citet{2013ApJS..209....5S} (v1) are marked in black. The energies, the fitted linear function and the power-law index estimated based on \citet{2007AN....328..904K} (v2) are marked in blue.} \label{fig:energie_cumulative}}
\fig{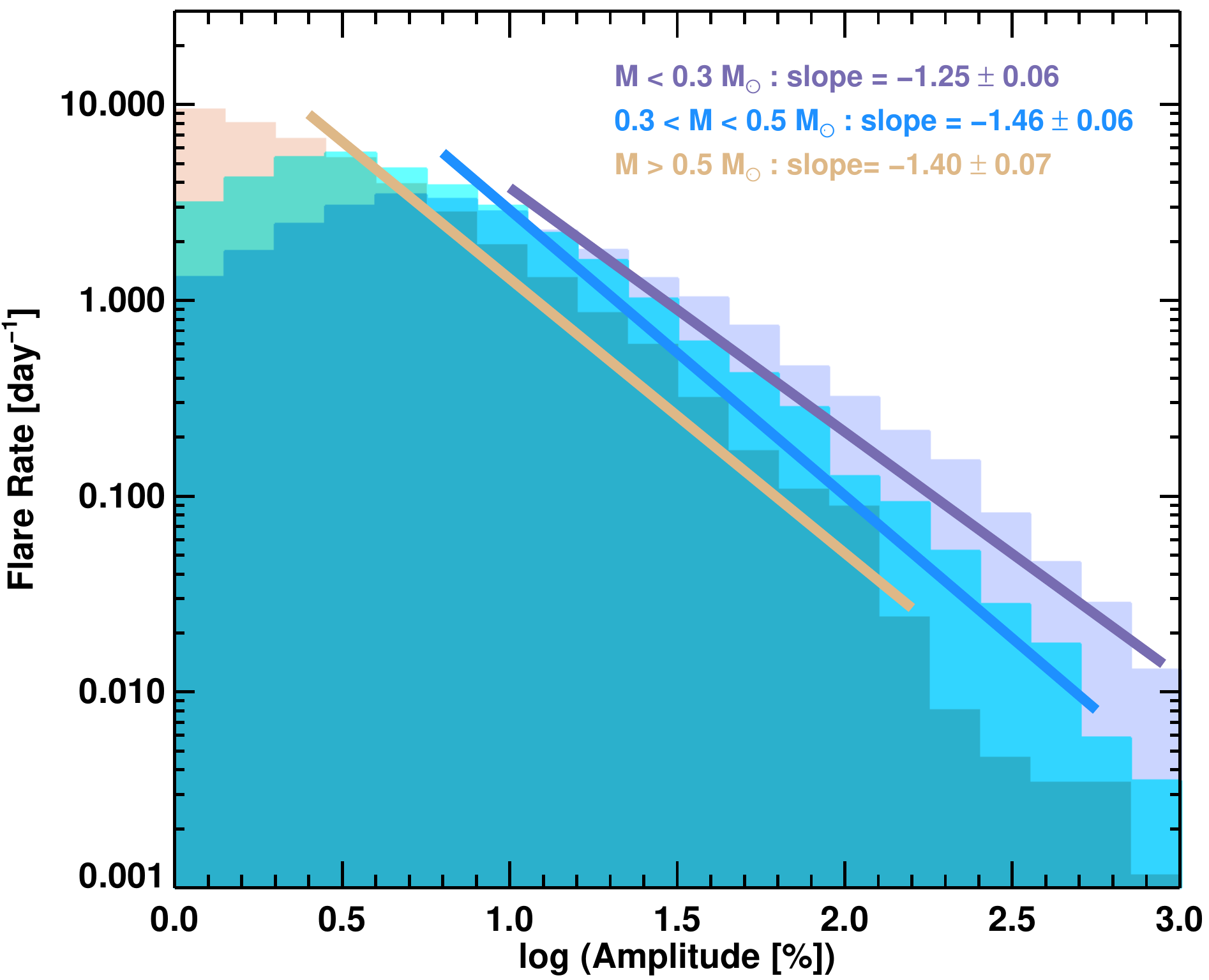}{0.427\textwidth}
{\caption{The flare frequency distributions (FFDs) of stars binned by the flare amplitude. The distributions were divided into three samples of stars. One of them was stars with masses less than 0.3 M$_\odot$ (violet), second one with masses in between 0.3 M$_\odot$ and 0.5 M$_\odot$ (blue), and the other with masses greater than 0.5 M$_\odot$ (beige). The lines show the slopes of the distributions. The slope value for each sample is given in the upper-right corner.} \label{fig:energie_mass}}
}
\end{figure}

Another result of the performed calculations is the estimation of energy of most stellar flares. Figure \ref{fig:energie_cumulative} shows the cumulative energy distribution estimated using both of the previously discussed methods based on \citet{2013ApJS..209....5S} (named v1, dark blue in Figures 15, 17, 20) and \citet{2007AN....328..904K} (named v2, light blue in Figures 15, 17, 20). Stars that have energies estimated by both methods must have such parameters as: radius, effective temperature, and \textit{log(g)}. The sample of flares that have energy estimated only by the first method is larger by about 3,000 flares, because the \textit{log(g)} is not necessary then. The index of power-law approximation is about 1.68 fort the method based on \citet{2013ApJS..209....5S} and about 1.65 for the method based on \citet{2007AN....328..904K} in the flare energy from $5\times10^{33}$ to $10^{36}$ erg. These results are in agreement with the previous papers \citep{1976ApJS...30...85L,2014ApJ...797..121H,maehara_2020}. 
The flare frequency distributions (FFDs) of stars binned by the flare amplitude are presented in Figure \ref{fig:energie_mass}. We examined how the slopes of the distributions vary for three samples. One of them was stars with masses less than 0.3 M$_\odot$ (violet), second one with masses in between 0.3 M$_\odot$ and 0.5 M$_\odot$ (blue), and the other with masses greater than 0.5 M$_\odot$ (beige). The slope values of the distributions are very similar to those in other works and is about -1.4 (\citep{2020JOSS....5.2347F,2022Fein}. It can be seen that a sample of stars with lower masses also has a shallower slope. The differences in slopes can be explained by theoretical works \citep{2022ApJ...929...54S}.

\begin{figure}[H]
\gridline{
\fig{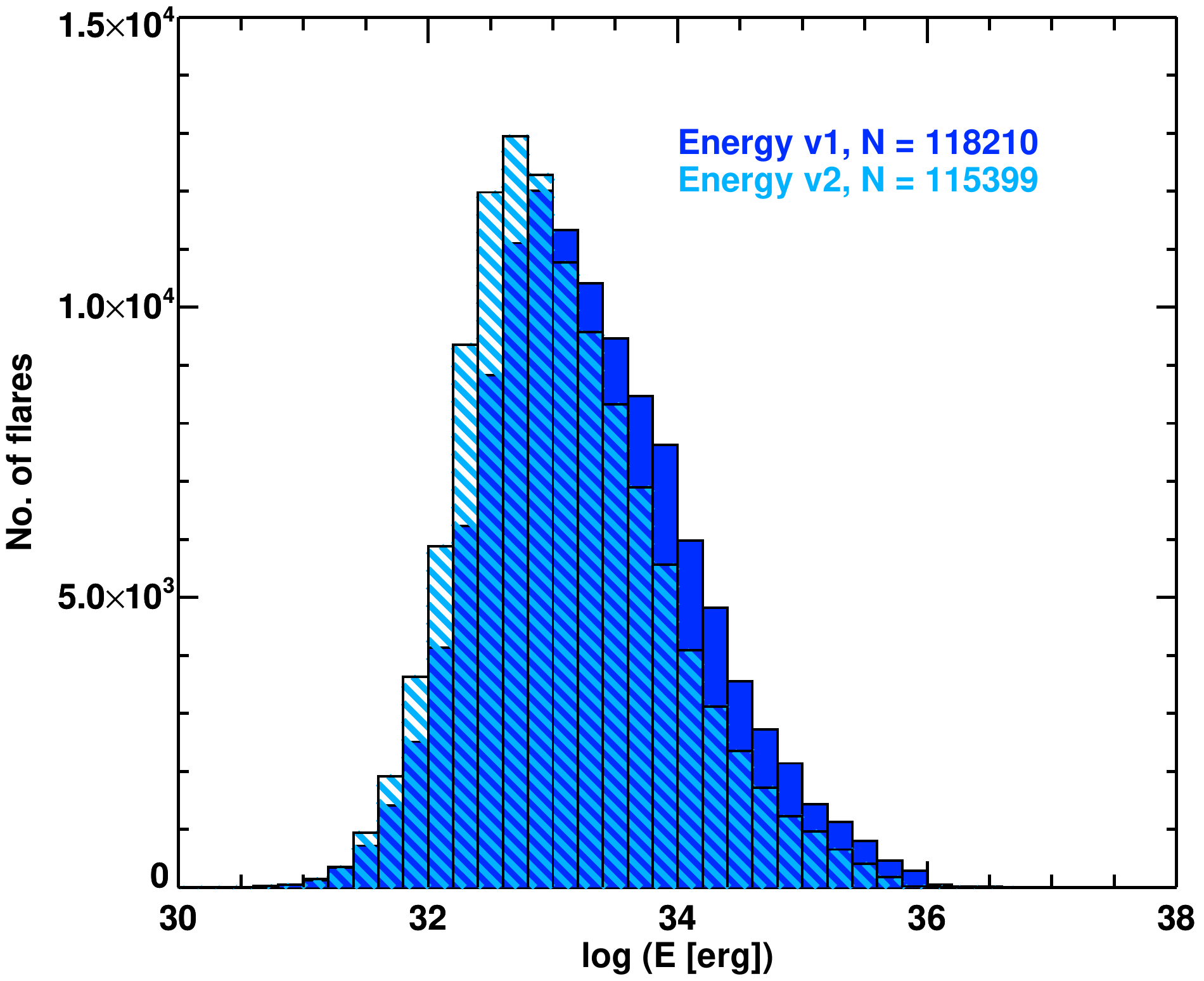}{0.42\textwidth}
{\caption{Distribution of flare energies. 
The energies estimated using the method based on \citet{2013ApJS..209....5S} (v1) are marked in dark blue and based on \citet{2007AN....328..904K} (v2) are marked in light blue.} \label{fig:energie1}}
\fig{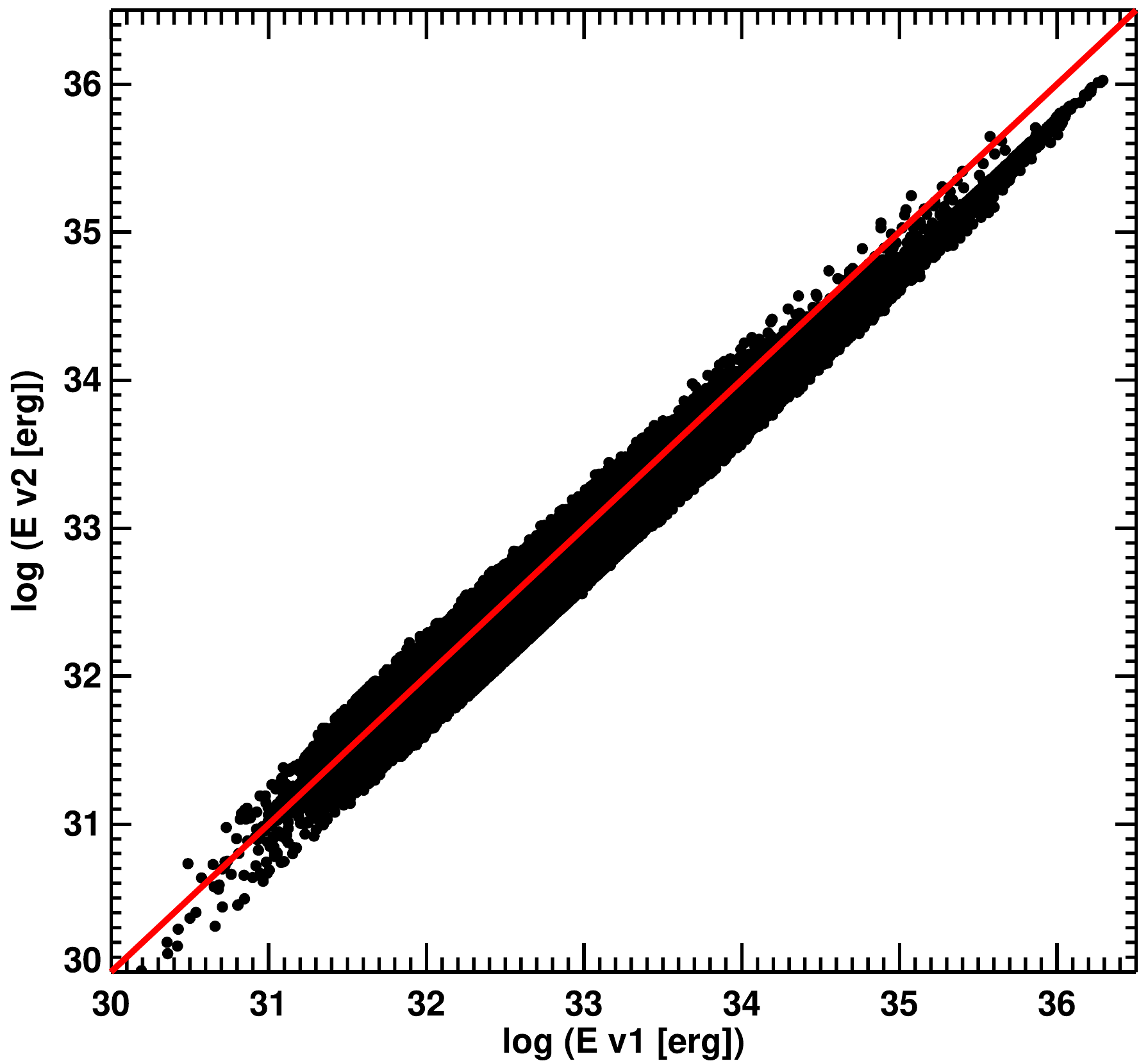}{0.368\textwidth}
{\caption{The dependence between flares' energies estimated using the method based on \citet{2013ApJS..209....5S} (v1) and based on \citet{2007AN....328..904K} (v2). The red line represents the same energies in both methods.}}\label{fig:ev1_ev2}
}
\end{figure}

\begin{figure}[H]
\gridline{
\fig{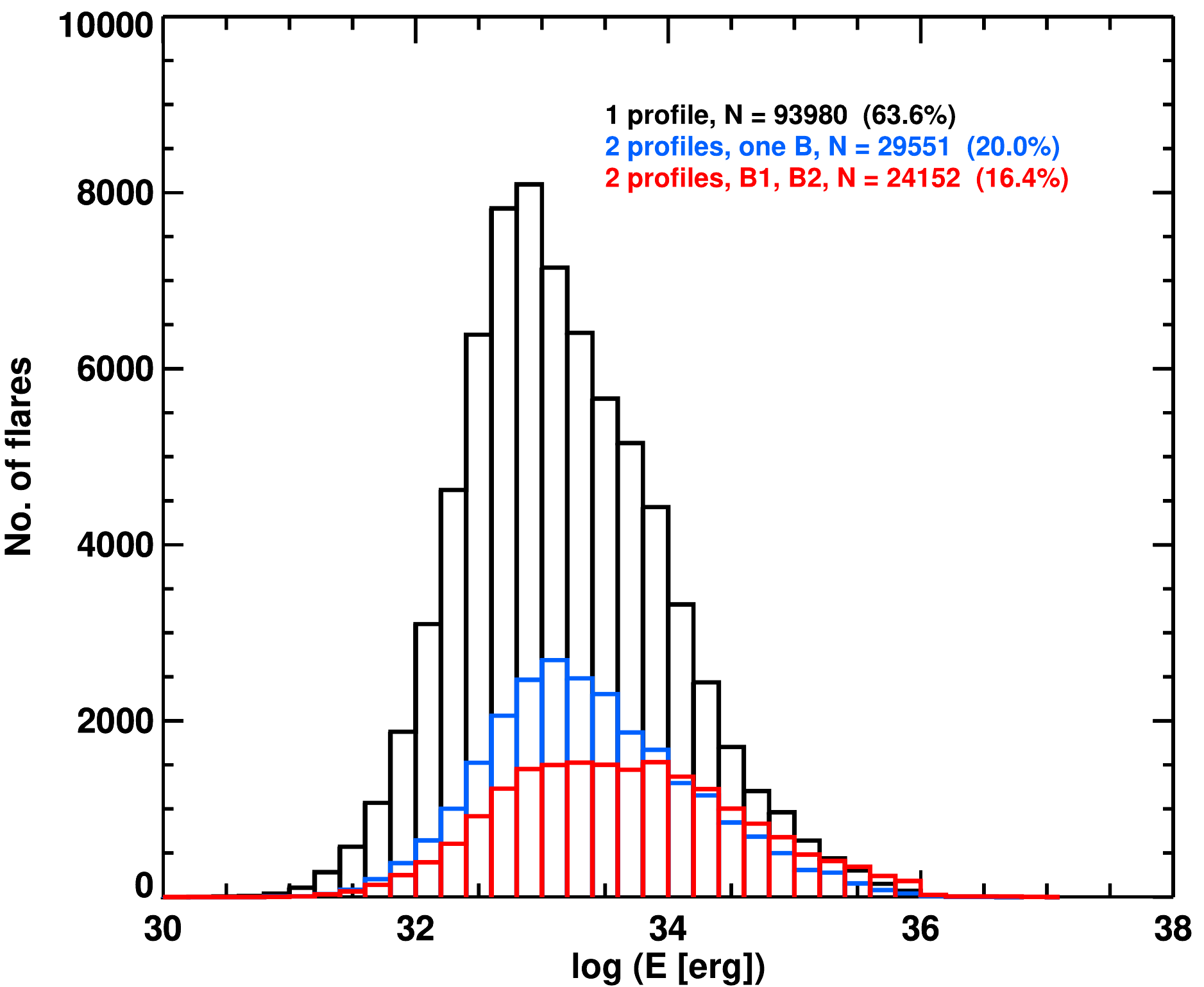}{0.415\textwidth}
{\caption{Distribution of the flare energies fitted with: one profile (black), two profiles with one B (blue), and two profiles with B1 and B2 (red).} \label{fig:energie_prof}}
\fig{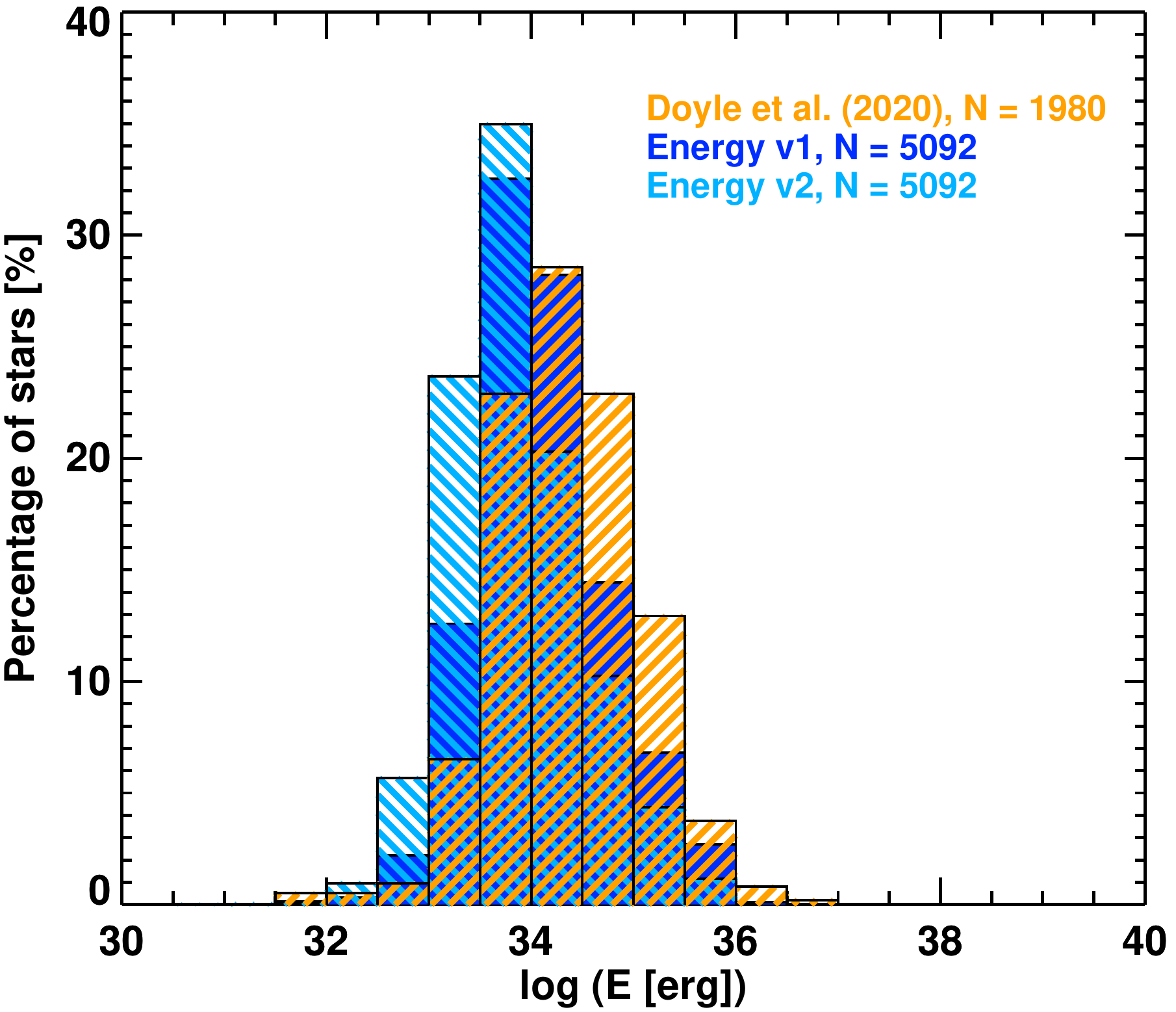}{0.4\textwidth}
{\caption{The distribution of the energies of the 1980 flares from \citet{2020MNRAS.494.3596D} and from our sample of stars from \textit{TESS} sectors 1-13 (S01-S13.)} \label{fig:energie2}}
}
\end{figure}

Figure \ref{fig:energie1} shows the histogram of flares' energies. It could be noticed that the results obtained with the first method \citep{2013ApJS..209....5S} are larger than those obtained with the second method \citep{2007AN....328..904K}. The estimated energies of stellar flares range from 10$^{31}$ to $5\times10^{36}$ erg, which means that most of the detected events are super-flares. Figure \ref{fig:ev1_ev2} additionally  shows the dependence between flares' energies estimated using the method based on \citet{2013ApJS..209....5S} (v1) and based on \citet{2007AN....328..904K} (v2). Our results were compared with the works: \citet{2020MNRAS.494.3596D},  \citet{gunther}, \citet{2019ApJ...881....9H}. These studies also investigated stellar flares observed by the \textit{TESS} satellite. The comparison shows that our results agree with the already known average energy of the detected stellar flares. Figure \ref{fig:energie_prof} shows the differences between the energy distribution of flares fitted with one or two profiles. Based on the Kolmogorov–Smirnov test we reject the hypothesis of the same distributions. The flares with the highest energies (greater than 10$^{35}$ erg) are usually fitted with two profiles.

\newpage

\begin{figure}[H]
\gridline{
\fig{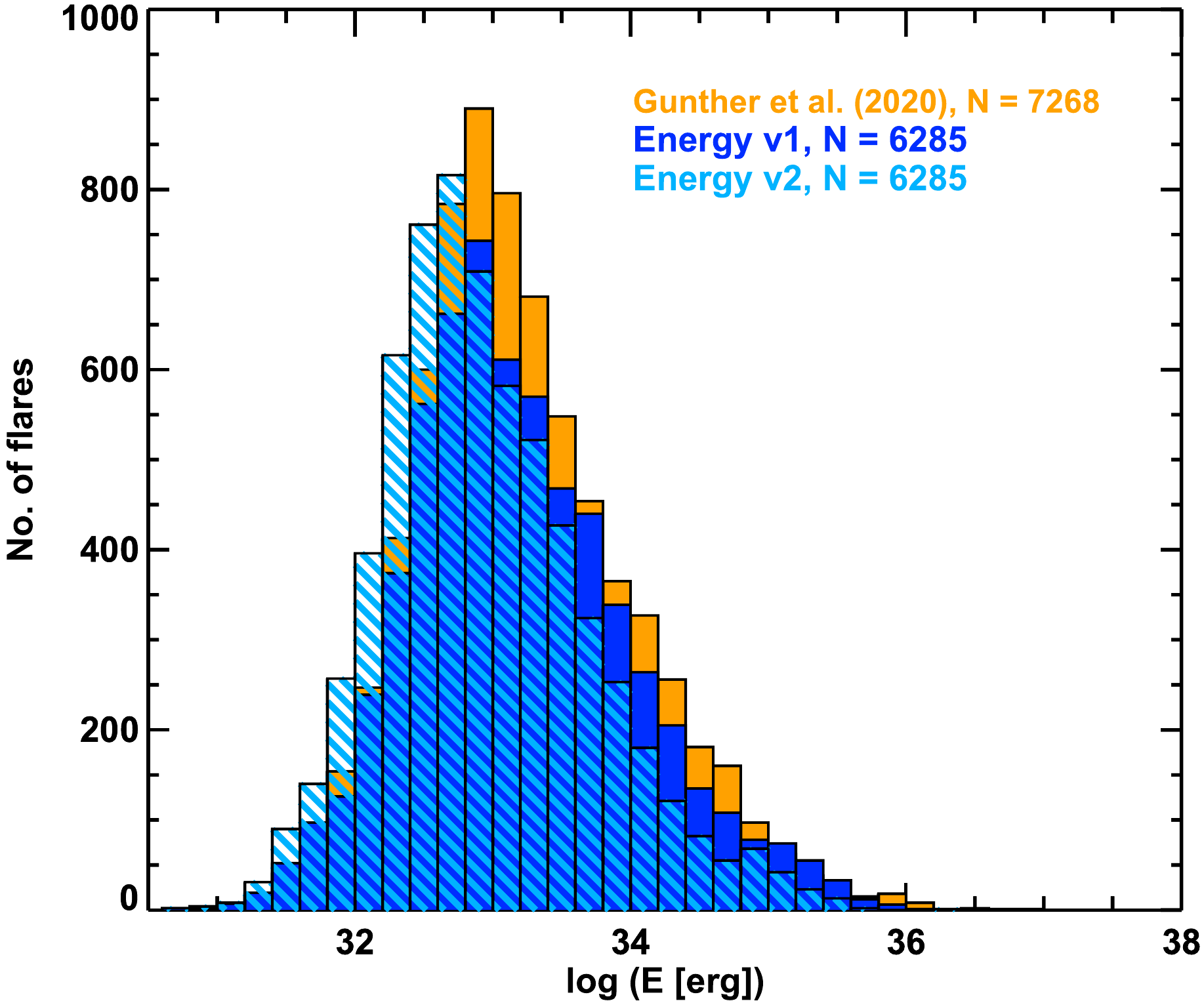}{0.415\textwidth}
{\caption{Distribution of the energies of the flares from \citet{gunther} and from our sample of stars from \textit{TESS} sectors S01-S02.} \label{fig:energie3}}
\fig{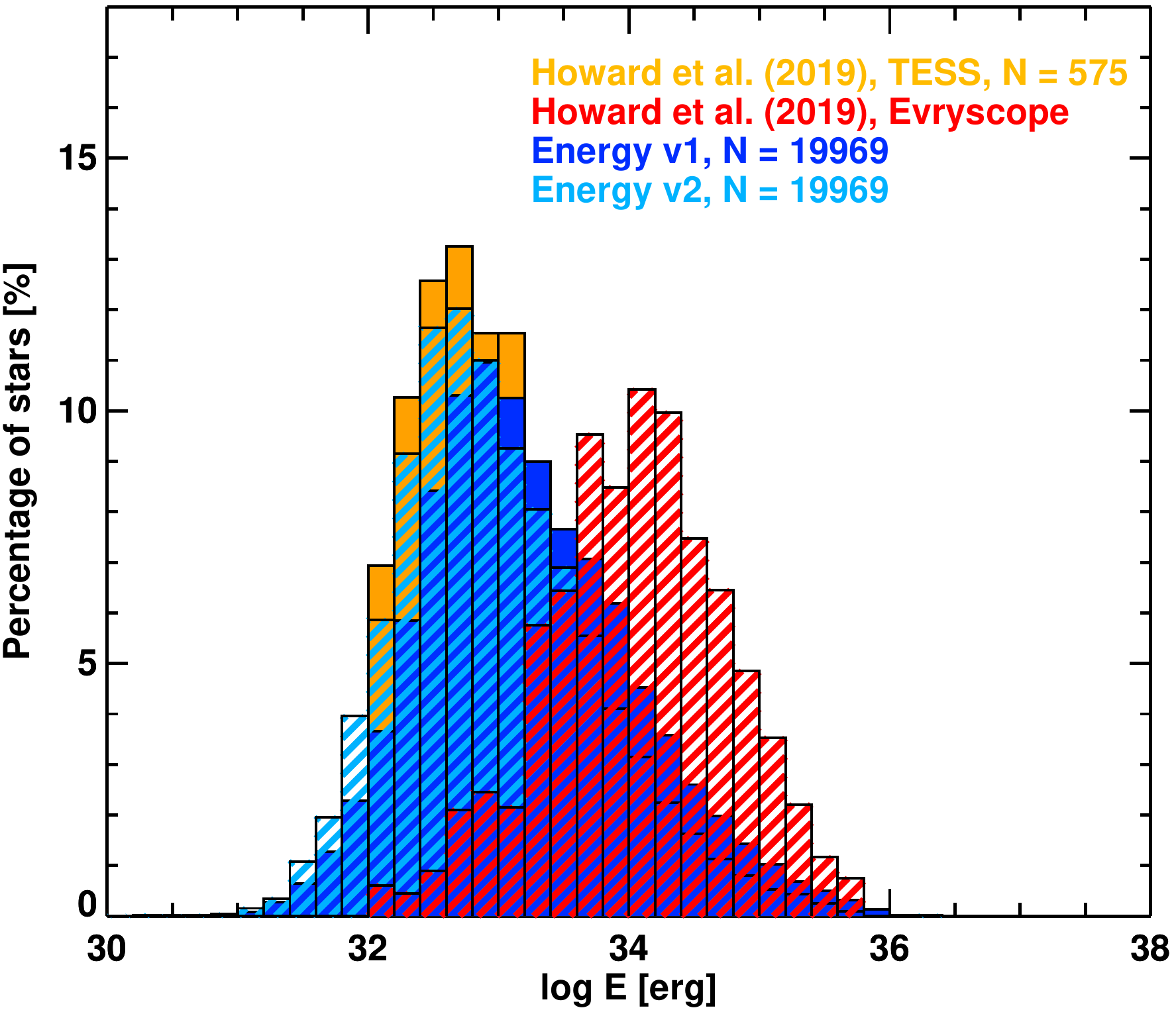}{0.4\textwidth}
{\caption{Normalized distribution of bolometric flare energies from Evryscope and \textit{TESS} sectors S01-S06. The results obtained by \citet{2019ApJ...881....9H} are marked in yellow and red.} \label{fig:energie4}}}
\end{figure}

Figure \ref{fig:energie2} shows the flare energies distribution calculated in our work and that from \citet{2020MNRAS.494.3596D}. The energy ranges from 10$^{32}$ to 10$^{36}$ erg and the maximum of the distribution from \citet{2020MNRAS.494.3596D} is for energies about four times higher than in our work. This is due to the fact that we have modified the method of estimating the energy of the flares described in \citet{2007AN....328..904K}. The flare energies obtained in our sample are very similar to those in \citet{gunther} (Figure \ref{fig:energie3}). Small differences could be attributed to different methods of searching for stellar flares used. Moreover, in \citet{gunther} flares of shorter duration (from six minutes) than in our work were included. Figure \ref{fig:energie4} shows another comparison. \citet{2019ApJ...881....9H} estimated flare energies based on the \textit{TESS} observations in sectors 1-6 and the Evryscope. The average energy of the flares from both datasets is significantly different. The authors explain that \textit{TESS} observed much fainter, lower amplitude events. Our estimations are similar to the energies obtained by other authors from the observations from \textit{TESS}. Based on all presented results, it can be seen that the energy of stellar flares varies from  $10^{31}$ erg to $10^{36}$ erg, depending on the analyzed star sample.

\begin{figure}[H]
\gridline{
\fig{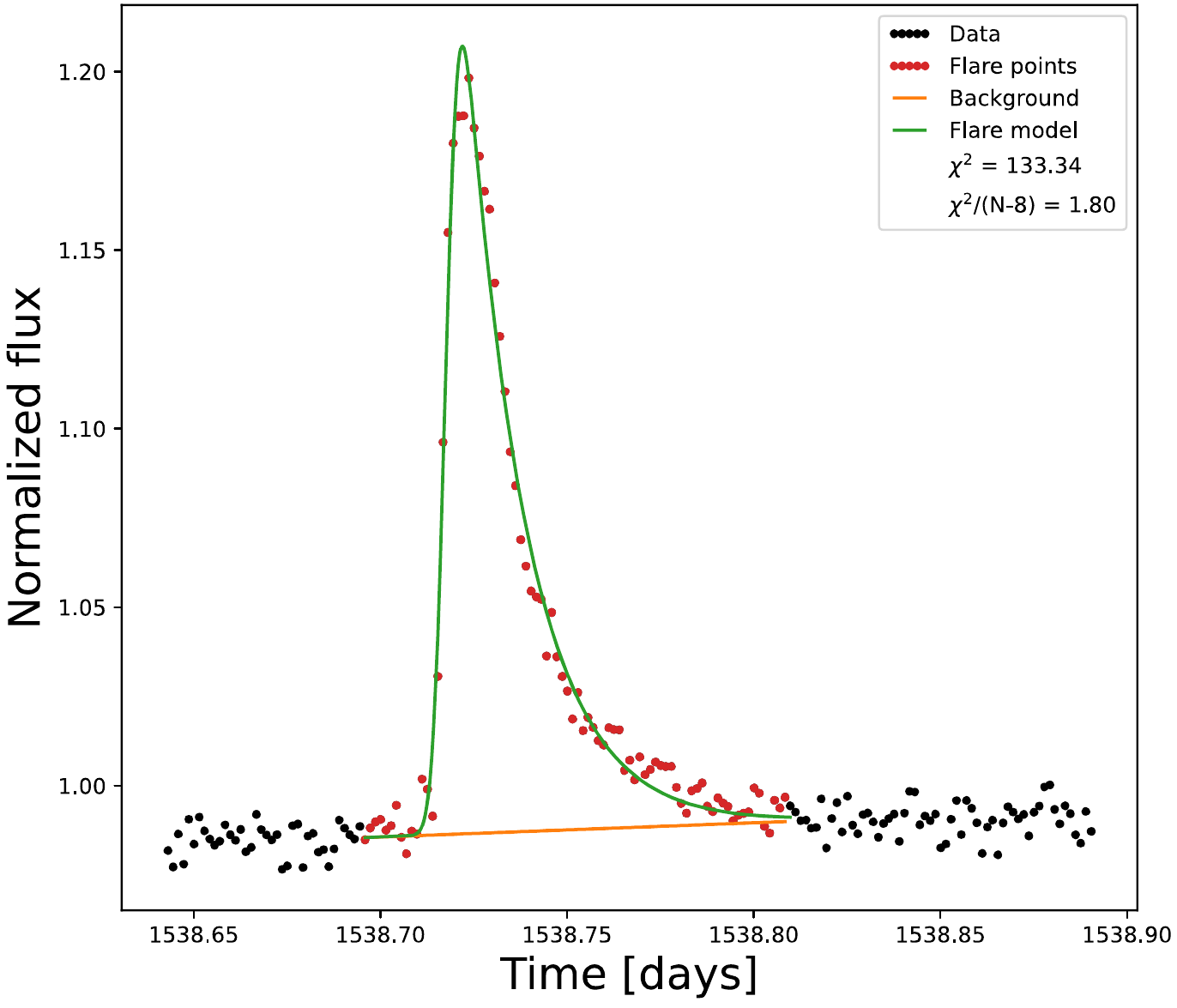}{0.43\textwidth}
{\caption{Super-flare detected on TIC175305230. The fitted profile is marked with a green line.} \label{fig:tic175305230}}
\fig{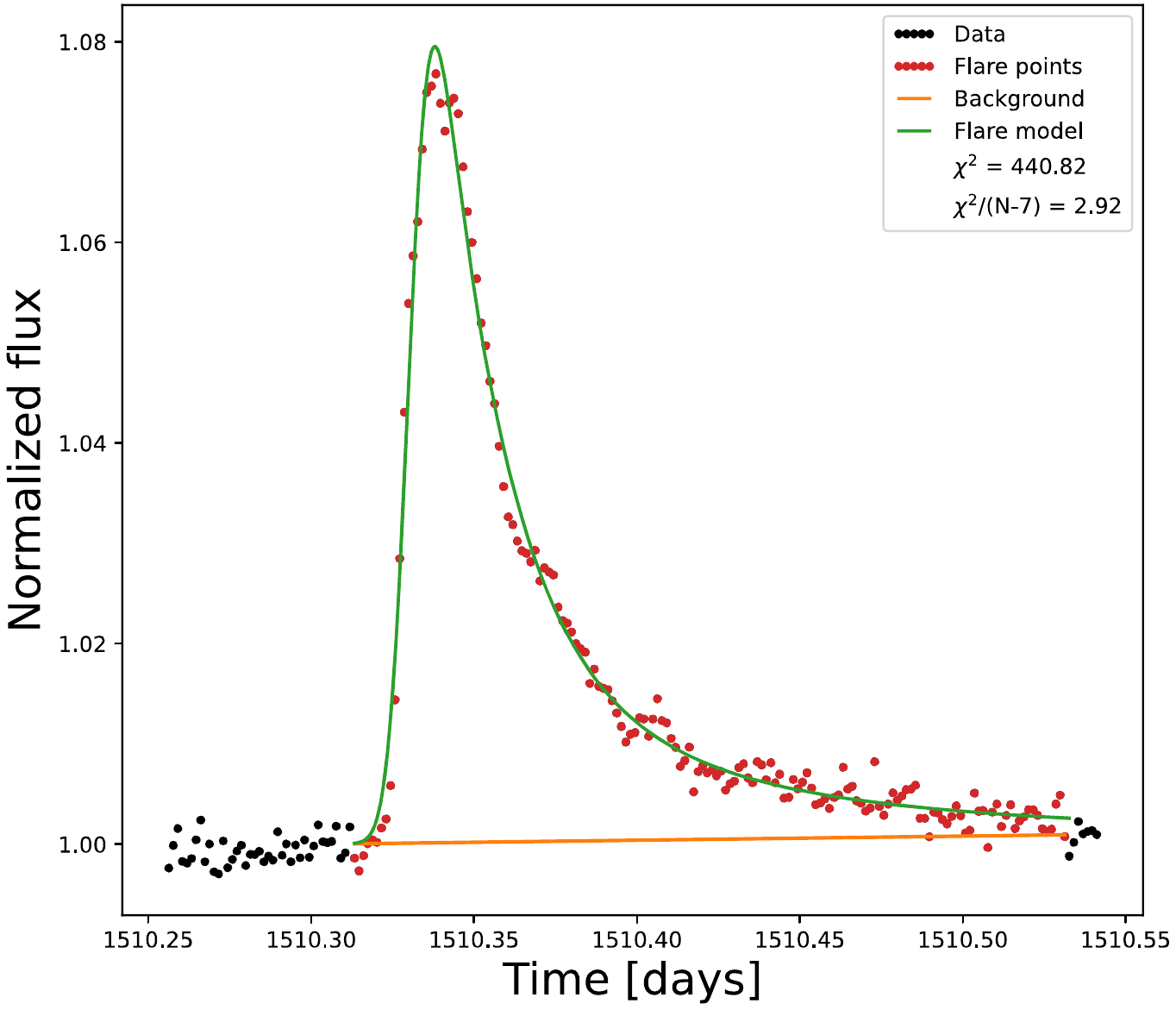}{0.43\textwidth}
{\caption{Super-flare detected on TIC232036408. The fitted profile is marked with a green line.} \label{fig:tic468191285}}}
\end{figure}

Table \ref{tab:table_energies} additionally summarizes information about the stars with the most energetic flares found by our software. Among the super-flares listed here, we found only one that was already mentioned in other works. We estimated the energy of the flare on the TIC79358659 at $10^{35.88}$ erg, and in \citet{2020MNRAS.494.3596D} at $10^{35.72}$ erg. We would like to notice the systematic overestimation of the flare energy is caused by too large estimates of the stellar radius from the MAST catalog. In the analyzed sample of stars, there were objects classified as main-sequence stars, but with a radius of several or a dozen solar radii. For this reason, we limited our analysis only to stars with a radius smaller than 2 R$_\odot$. Examples of super-flares on  TIC175305230 and TIC232036408 stars with fitted profiles are shown in Figures \ref{fig:tic175305230} and \ref{fig:tic468191285}. The flares were fitted with two profiles with one B. Assuming that the bolometric energy of 10$^{33}$ erg corresponds to an SXR scale X100 flare \citep{article}, the detected events are more energetic than X100000 flares.

\begin{deluxetable}{ccccccccc}[H]
\tablenum{4}
\tablecaption{The stars with the flares of the highest energies
\label{tab:table_energies}}
\tablewidth{0pt}
\tablehead{
\colhead{ID} & \colhead{Name} &  \colhead{Energy v1} &  \colhead{Energy v2} & \colhead{Duration}  & \colhead{Amplitude}  &  \colhead{Spectral type} &  
\colhead{T$_{\rm{eff}}$ }& \colhead{Radius} \\
\colhead{} & \colhead{} & \colhead{[erg]} & \colhead{[erg]}  & \colhead{[min]} & \colhead{[rel. flux]} & \colhead{} &
\colhead{[K]}  & \colhead{[R$_\odot$]}} 
\startdata
TIC464498986      &TYC 8604-1093-1 & 3.05e+36  & 1.47e+36  &  336  & 0.515 & K3V\tablenotemark{b}      & 4869     & 1.1     \\ 
TIC79358659     & HD 321958 & 9.88e+35   & 5.92e+35   & 184 &  0.154   & G9V\tablenotemark{a}     & 5626         &  1.3     \\
TIC175305230      & Cl* NGC 2451 AR 52 & 9.07e+35  & 5.30e+35   & 170 & 0.228  & G8V\tablenotemark{b}      & 5419     & 1.1     \\ 
TIC468191285      &  LEHPM 4131  & 8.43e+35  & 4.70e+35     & 192 &  0.349    & G9Ve\tablenotemark{a}         & 4966        & 0.9     \\ 
TIC420137030     & TYC 4456-307-1 & 8.05e+35     & 4.65e+35  & 212 &  0.146   & G0V\tablenotemark{b} & 5935    & 1.1     \\
TIC50656379     & 2MASS J05292131-0028357 & 7.75e+35 & 3.88e+35      & 204 &    0.695   & M1V\tablenotemark{b}    & 3681        & 1.0     \\ 
TIC79358659     & HD 321958 & 7.65e+35   & 4.60e+35    & 216 &   0.103  & G9V\tablenotemark{a}     & 5626         &  1.3     \\
TIC50897755     & CVSO 119 & 6.88e+35     & 2.83e+35   & 186 &   0.311   & K7\tablenotemark{a}          & 4045        & 1.2     \\ 
TIC81100117     & 2MASS J08104428-4727558   & 6.15e+35  & 3.63e+35    & 198 &   0.256  & K3V\tablenotemark{b}   & 4885        & 0.8     \\ 
TIC232036408     & TYC 7621-604-1  & 5.85e+35  & 3.03e+35    & 220 &     0.077   & G3V\tablenotemark{b}          & 5717        & 1.1     \\ 
\enddata
\tablenotetext{\tiny a}{  SIMBAD}
\tablenotetext{\tiny b}{  \citet{2013ApJS..208....9P}}
\end{deluxetable}

\begin{figure}[H]
\fig{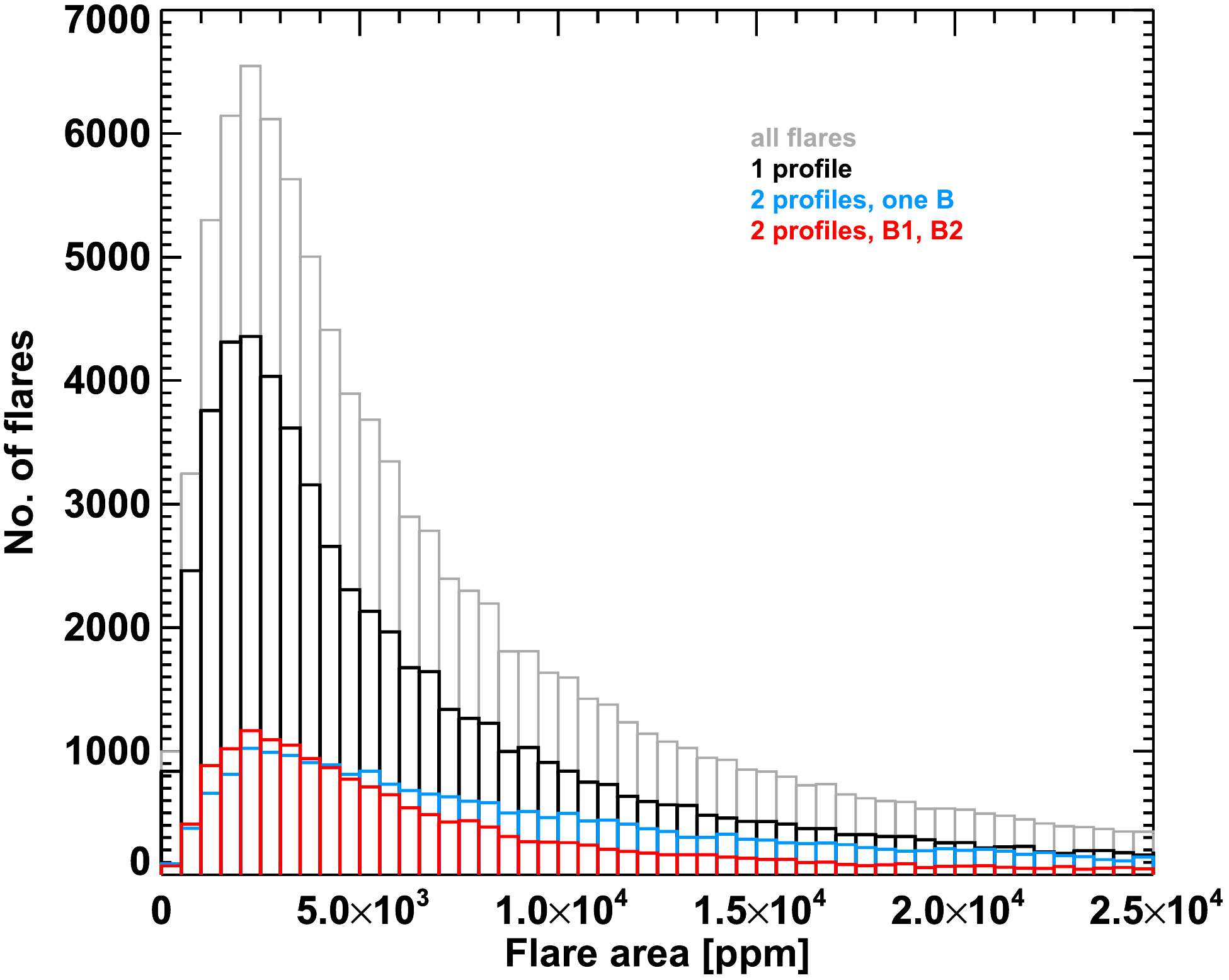}{0.6\textwidth}
{\caption{Distribution of the areas of the flares in parts per million of a stellar disc [ppm]. All flares are marked with gray, flares fitted with one profile are marked in black, two profiles with one B with blue, and two profiles with B1 and B2 with red.} \label{fig:area2}}
\end{figure}

Using the energy estimation method proposed in \citet{2013ApJS..209....5S}, it is also possible to determine the areas of stellar flares (Equation \ref{eq:area}). Figure \ref{fig:area2} shows a distribution of the areas of the flares obtained in our research in parts per million of a stellar disc [ppm]. All flares are marked with gray, flares fitted with one profile with black, two profiles with one B with blue, and two profiles with B1 and B2 with red. The maximum of each of the distributions is about 2200 ppm. Based on the Kolmogorov–Smirnov test we rejected the hypothesis of the same distributions. 
For the fit with one profile, the average value of the flares areas is 0.31\% of the entire stellar disk, and for the fits with one and two B parameters it is 0.35\% and 0.27\% respectively. Flares fitted with two profiles with one B parameter are characterized by on average larger areas than other flares. Our results are in line with the general knowledge of the areas of solar and stellar flares observed in white light \cite{Heinzel2018CanFL}. 

\begin{figure}[H]
\gridline{
\fig{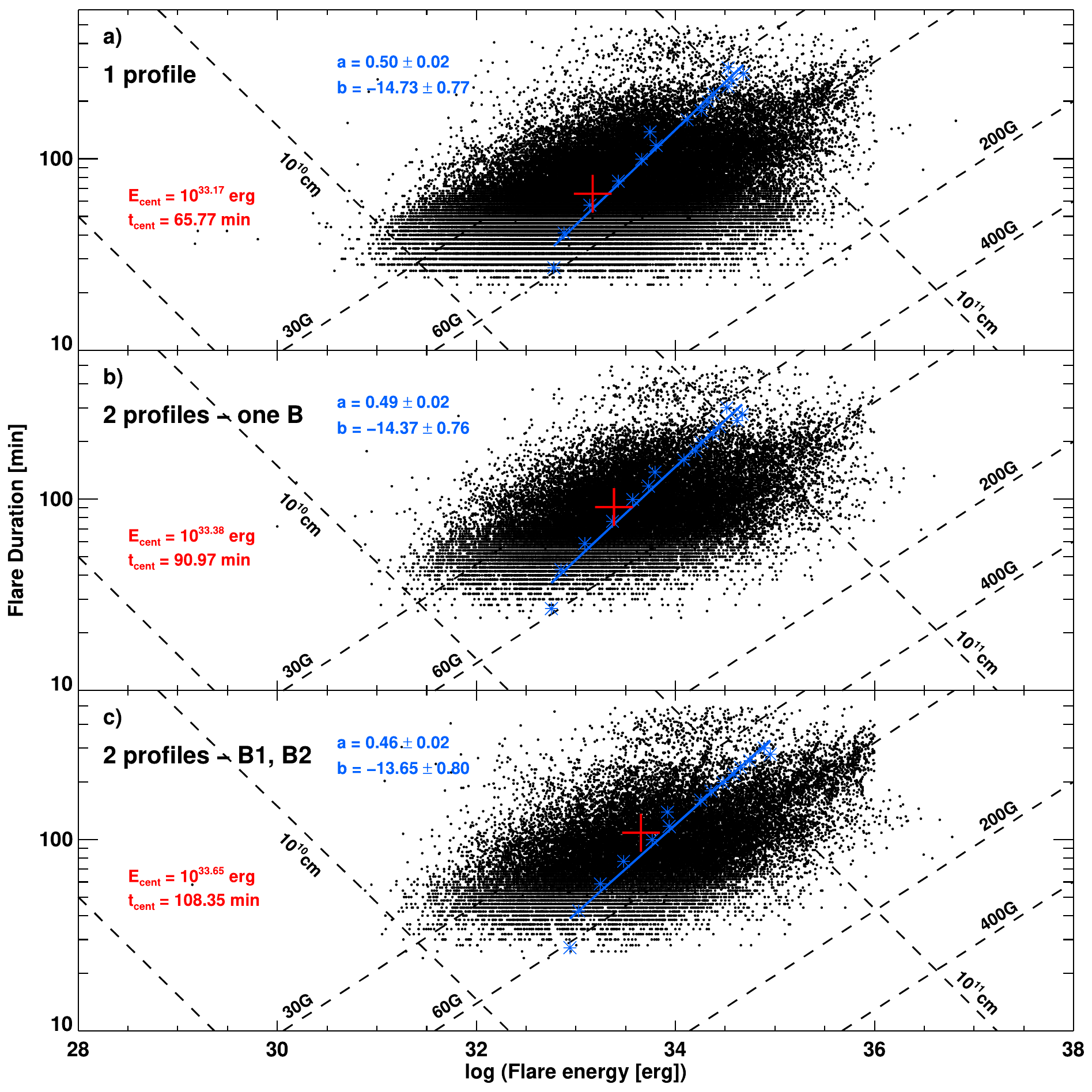}{0.9\textwidth}
{\caption{Observed relations between flare energy and duration with theoretical dotted lines for the different strength of magnetic fields and the lengths of the flare loops. A constant pre-flare coronal density is assumed. On the panel a) there are marked flares fitted with one profile, on the panel  b) flares fitted with two profiles with one B and on c) fitted with two profiles with B1 and B2. Discrete flare durations bins result from the time resolution of observations equal to 2 minutes. In each case, the red color shows the location of the center of mass with its coordinates. The blue color shows a linear fit to the data. More information is in the text.} \label{fig:namekata}}
}\end{figure}

An analysis similar to that in \citet{2017ApJ...851...91N} or \citet{2021ApJS..253...35T} was also carried out in our work. These authors compare the durations and energies of the flares observed on the Sun and other solar-like stars. Based on the scaling laws, they estimated the magnetic field strengths and the lengths of the flares loops. The authors assume a constant pre-flare coronal density. Other versions of the scaling laws, in which the coronal density changes with the length of the flare loop, have also been proposed. Moreover, the coronal density of very active stars may be much greater than the solar coronal density. This can be concluded from their short rotation periods and the emission in the X-ray range. Therefore, the obtained results are only estimates of the strength of magnetic fields and the lengths of the flare loops. Scaling laws used in this work are given by the following equations \citep{2017ApJ...851...91N}: 

\begin{equation}
\tau \propto E^{1/3} B^{-5/3}
\end{equation}

\begin{equation}
\tau \propto E^{-1/2} L^{5/2}
\end{equation}

where $\tau$ is the flare duration time, $E$ is the estimated flare energy, $B$ is the magnetic field strength, and $L$ is the length of the flare loop. Figure \ref{fig:namekata} shows the observed relations between flare energy and duration from our sample of stars. On this graph, we have plotted theoretical lines for the different strength of magnetic fields (30, 60, 200, 400 G) and the lenghts of the flare loops (10$^{10}$ cm, 10$^{11}$ cm). The upper panel shows flares fitted with one profile, the middle one flares fitted with two profiles with one B, and the lower panel flares fitted with two profiles with B1 and B2. The location of the center of mass with its coordinates is marked with red. The centers of mass shifted towards greater flare energies and longer durations. Moreover, the crosses marked in blue mean the average energy of the flares in a given range of their duration (10-30 minutes, 30-50 minutes, etc.). The linear fit to these points with their parameters is also marked in blue. The slope of the line changes from 0.49 to 0.46 for the flares fitted with different profiles.

For flares fitted with two profiles, the dependence between the flare energy and the duration is stronger. Flares for which it is necessary to distinguish two components of the profile are usually stronger, longer, and more complicated.
We assume that there are two mechanisms in such flares: direct heating of the photosphere with non-thermal electrons and back-warming processes. We assume that there are two mechanisms in such flares: direct heating of the photosphere with non-thermal electrons and back-warming processes which are of thermal origin. For example, observations of flares on the AU Mic star with the FUSE (Far Ultraviolet Spectroscopic Explores) satellite shows an increasing continuum flux to shorter wavelengths, which can be interpreted as free-free emission for hot plasma \citep{Redfield_2002}. Our results are similar to those from \citet{2017ApJ...851...91N}. The estimated energies and the durations of the flares agree very well with the analogous values for the super-flares observed by \textit{Kepler} with a 30-minute cadence. The high values of magnetic field strengths up to 400 G determined by the scaling laws are characteristic for shorter super-flares observed by \textit{Kepler} with a cadence of 1 minute.  

\begin{figure}[H]
\gridline{
\fig{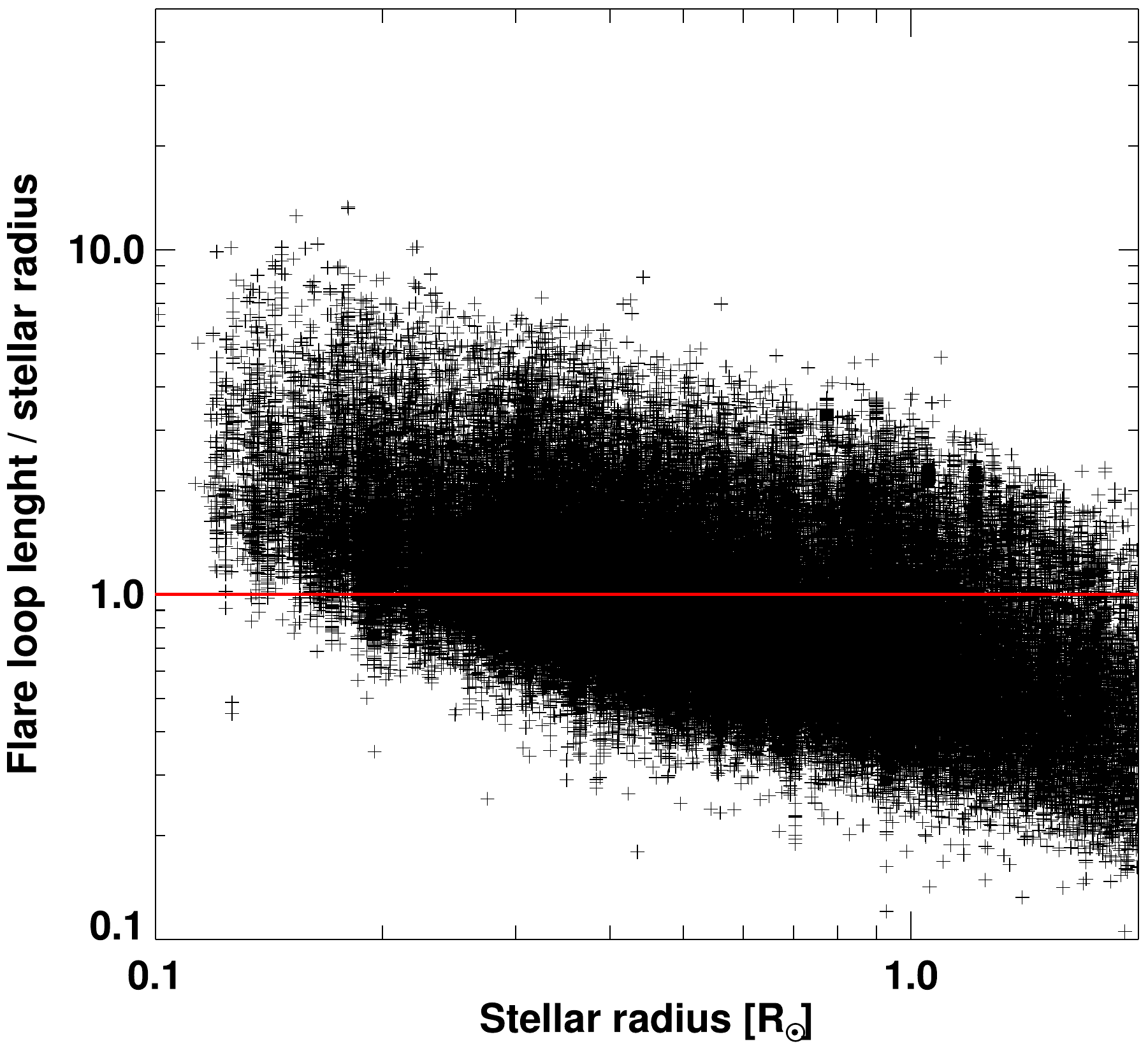}{0.55\textwidth}
{\caption{The relationship between the estimated flare loop length and stellar radius.} \label{fig:petle_radius}}
}
\end{figure}

We assume a constant pre-flare coronal density. Using information about the flares' durations and its energies we calculated magnetic field strengths and flare loop lengths. The average values of the magnetic field strengths are from 10 G to 200 G. The estimated lengths of the flare loops are from 10$^{10}$ cm to $2\times10^{11}$ cm. These values correspond to even several radii of a given star (Figure \ref{fig:petle_radius}). It could be concluded that the single-loop assumption is not sufficient for most of the observed flares. Stellar flares are probably associated with several or more flare loops. This conclusion should be correct taking into account the observations of strong events occurring on the Sun \citep{aschwanden}.

\section{Discussions}  \label{sec:discussion}
We analyzed the first 39 sectors of \textit{TESS} satellite data (about 330,000 stars). More than 140,000 flares were detected on over 25,000 stars. The flaring stars that we found make up about 7.7\% of the entire data sample. This value varies greatly depending on the effective temperature or the spectral type of a star. Most of found flaring stars are in the main-sequence, but we detected flares on young stars and giants as well. The effective temperatures for most of flaring stars do not exceed 8,000 K. The percentage of flaring stars could be even greater than 50\% for the stars of spectral type M. 

In total, during the first year of \textit{TESS} observations (S01-S13, ecliptic declination from -90$^\circ$ to -54$^\circ$), we found 9480 flaring stars. For comparison, during the third year of the mission, when the southern hemisphere was re-observed (S27-S39), we found 10661 flaring stars, including 8913 new objects.

In our work, we used three types of flare profiles: single, double with one B, and double with B1 and B2. A single profile was sufficient for more than half of the events. The maximum of the all flares duration distribution is approximately 50 minutes. The flares fitted with two profiles are characterized by an average longer duration. The average value of the durations of flares fitted with one profile is 66 minutes, and with two profiles is 98 minutes. The longest observed events last up to several hours. The average values of the growth times are much shorter than the decay times and are usually below 10 minutes.

Usually longer and stronger events are fitted with two profiles. One of these profiles is supposed to be related to the direct heating of the photosphere by non-thermal electrons. It is characterized by very short growth and decay times (parameters B, C and D). The second one, longer (with significantly lower parameter D) may be the result of back-warming processes \citep{2007Isobe}. In both double profiles, the second component dominates, especially in the decay phase.

We also estimated the energies of stellar flares based on two different methods (\citet{2013ApJS..209....5S} and modified \citet{2007AN....328..904K}). In the first method, we assumed a flare temperature of 10000 K, while in the second method no such assumption was needed. In most cases, the energy calculated using the stellar spectrum gives lower values. This is due to the fact that the energy estimated in this way is not bolometric energy. Estimated energies of stellar flares range from 10$^{31}$ to 10$^{36}$ erg, which means that most of the detected events are super-flares. Our results were compared with the works: \citet{2020MNRAS.494.3596D},  \citet{gunther}, \citet{2019ApJ...881....9H}. Obtained flare energies are very similar in our sample and \citet{gunther} (Figure \ref{fig:energie3}). Small differences could be caused by different methods of searching for stellar flares. Moreover, in \citet{gunther} flares of shorter duration (from six  minutes) than in our work were included. Our estimations are similar to the energies obtained from \textit{TESS} observations by other authors. The index of power-law approximation at the cumulative distribution of the flare energies (Figure \ref{fig:energie_cumulative}) is about 1.7 for the method based on \citet{2013ApJS..209....5S} and about 1.5 for the method based on \citet{2007AN....328..904K} in the flare energy from $5\times10^{33}$ to $10^{36}$ erg. These results are in agreement with previous papers \citep{1976ApJS...30...85L,2014ApJ...797..121H,maehara_2020}.

Using the energy estimation method from \cite{2013ApJS..209....5S}, it is also possible to determine the area of stellar flares. The maximum of this distribution for all flares is about 2200 ppm (0.22\%) (Figure \ref{fig:area2}). Flares fitted with two profiles with one B parameter are characterized by on average larger areas than other flares. For the fit with one profile, the average value of the flares' areas is 0.31\% of the entire stellar disk and for fits with one and two B parameters it is 0.35\% and 0.27\% respectively. Our results are in line with the general knowledge of the areas of solar and stellar flares observed in white light \cite{Heinzel2018CanFL}.

The index of power-law approximation of the relation between flare energy and duration changes from 0.49 to 0.46 for the flares fitted with different profiles (Figure \ref{fig:namekata}). For the flares fitted with two profiles, the dependence between the flare energy and the duration is stronger. Flares for which it is necessary to distinguish two components of the profile are usually stronger, longer, and more complicated. Our results are similar to those from \citet{2017ApJ...851...91N}. The estimated energies and the duration of the flares agree very well with the analogous values for the super-flares observed by \textit{Kepler} with a 30-minute cadence. The high values of magnetic field strengths up to 400 G determined by the scaling laws are characteristic for shorter super-flares observed by \textit{Kepler} with a cadence of 1 minute. On the basis of the relationship between the length of the flare loops and the radius of the star (Figure \ref{fig:petle_radius}), it could be concluded that the single-loop assumption is not sufficient for most of the observed flares. It looks as if stellar flares are probably associated with several or more flare loops. This conclusion should be correct taking into account the observations of events occurring on the Sun.

\begin{figure}[H]
\gridline{
\fig{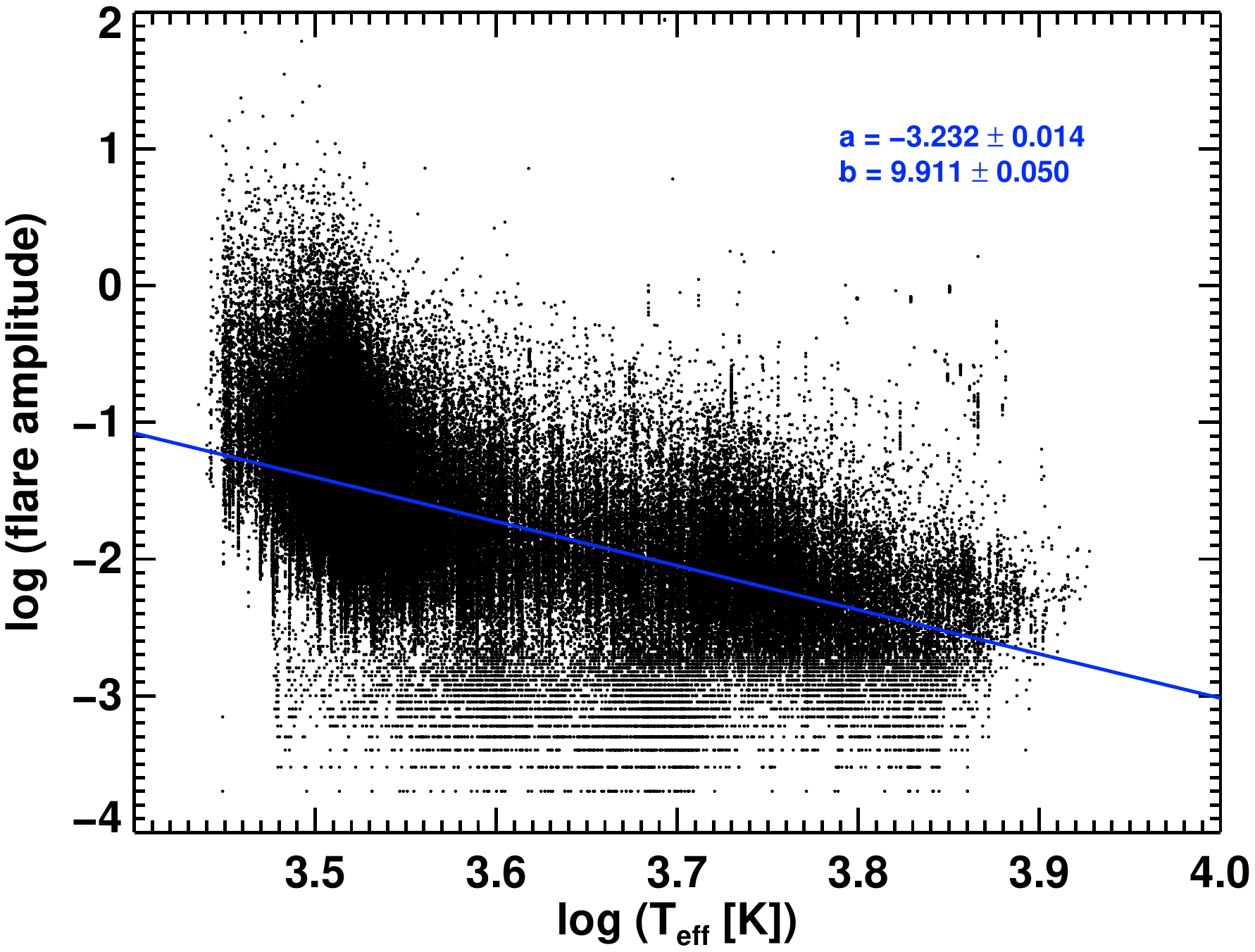}{0.55\textwidth}
{\caption{The relation between the logarithm of the amplitude of the flares and the logarithm of the effective temperature of the stars. The linear fit and its parameters are marked in blue.} \label{fig:teff_amp}}}
\end{figure}

Flares detected on cool stars have, on average, higher amplitudes due to the greater contrast between the flares' and the star's surface. Such an effect has already been noted in other studies analyzing stellar flares \citep{10.1093/mnras/stab979}. Figure \ref{fig:teff_amp} shows the relation between the logarithm of the amplitude of the flares and the logarithm of the effective temperature of the stars. The linear fit to this data and the linear model parameters are shown in blue. The linear model parameters are shown in the upper right corner. The big amplitude of the flare is not always related to the high energy emitted during its duration. For this reason, it is important to use methods to estimate the energy of stellar flares. Our results are consistent with previous studies on the topic based on the observations of \textit{TESS} and \textit{Kepler}.

An analysis of white-light stellar flares is important in understanding heating and emission processes during stellar flares. Since the mechanisms of continuum emission can be different from the mechanisms of spectral lines formation, observation of white-light flares can give us an additional constraint, allowing more precise spatial and temporal localization of heating, cooling, and energy transportation processes.

\textit{Acknowledgments.} This work was partially supported by the program "Excellence Initiative - Research University" for years 2020-2026 for University of Wrocław, project no. BPIDUB.4610.96.2021.KG.

The authors acknowledge also the \textit{TESS} consortium for providing the excellent observational data. This paper includes data collected by the \textit{TESS} mission, which are publicly available from the Mikulski Archive for Space Telescopes. Funding for the \textit{TESS} mission is provided by NASA’s Science Mission directorate.

The authors are grateful to an anonymous referee for constructive comments and suggestions, which have proved to be very helpful in improving the manuscript.

We would like to acknowledge the useful discussions with Professor P. Heinzel.

Facilities: \textit{TESS}.

Software: WARPFINDER \\


\textbf{ORCID iDs}

\noindent Małgorzata Pietras \href{https://orcid.org/0000-0002-8581-9386}{\includegraphics[height=0.3cm]{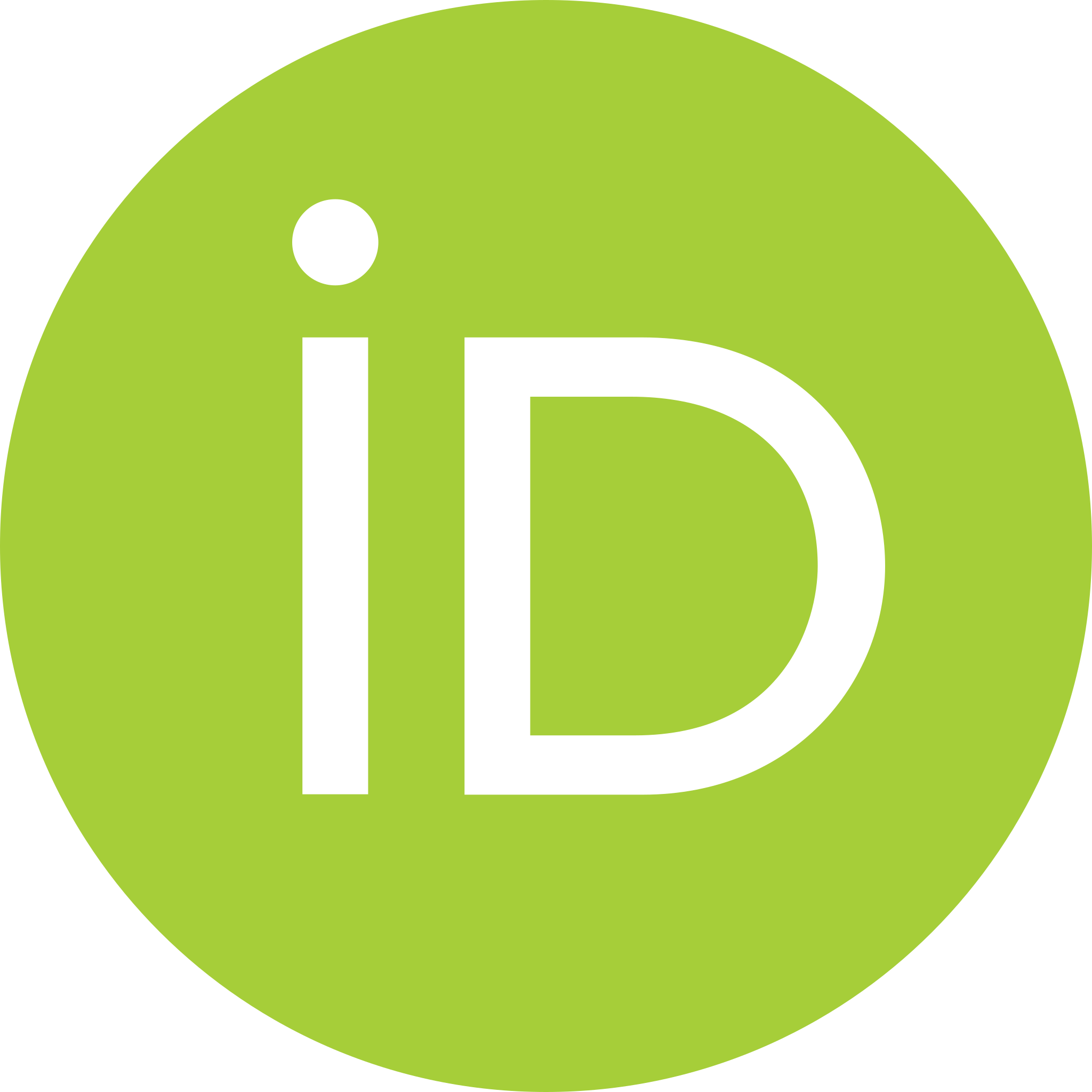}} \href{https://orcid.org/0000-0002-8581-9386}{https://orcid.org/0000-0002-8581-9386} \\
Robert Falewicz \href{https://orcid.org/0000-0003-1853-2809}{\includegraphics[height=0.3cm]{ORCID_iD.svg.png}} \href{https://orcid.org/0000-0003-1853-2809}{https://orcid.org/0000-0003-1853-2809} \\
Marek Siarkowski \href{https://orcid.org/0000-0002-5006-5238}{\includegraphics[height=0.3cm]{ORCID_iD.svg.png}} \href{https://orcid.org/0000-0002-5006-5238}{https://orcid.org/0000-0002-5006-5238} \\
Kmail Bicz \href{https://orcid.org/0000-0003-1419-28359}{\includegraphics[height=0.3cm]{ORCID_iD.svg.png}} \href{https://orcid.org/0000-0003-1419-2835}{https://orcid.org/0000-0003-1419-2835} \\
Paweł Pre\'s \href{https://orcid.org/0000-0001-8474-7694}{\includegraphics[height=0.3cm]{ORCID_iD.svg.png}} \href{https://orcid.org/0000-0001-8474-7694}{https://orcid.org/0000-0001-8474-7694} 

\bibliography{sample631}{}
\bibliographystyle{aasjournal}



\end{document}